\documentclass[sigconf, nonacm, pdfa]{acmart}
\usepackage{array}
\usepackage{multirow}
\usepackage{tikz}
\usepackage{caption}
\usepackage{subfigure}
\usepackage{colortbl}
\usepackage{graphicx}
\usepackage{makecell}
\usepackage{subcaption}
\usepackage[a-2b]{pdfx}
\theoremstyle{definition}
\newtheorem{definition}{Definition}[section]

\theoremstyle{plain}
\newtheorem{lemma}{Lemma}[section]
\usepackage{enumitem}

\usepackage[linesnumbered,ruled,vlined]{algorithm2e} 
\usepackage{algorithm2e,setspace}

\makeatletter
\renewcommand\subsubsection{\@startsection{subsubsection}{3}{\z@}%
  {-3.25ex\@plus -1ex \@minus -.2ex}%
  {1.5ex \@plus .2ex}%
  {\normalfont\normalsize\bfseries}}
\makeatother

\newcommand{\opentopsquare}{%
  \tikz{\draw[line width=0.65pt] (0,1.2ex) -- (0,0) -- (1.2ex,0) -- (1.2ex,1.2ex);} 
}
\newcommand{\customtriangle}{%
  \tikz{\draw[line width=0.65pt] (0,0) -- (0.6ex,1.2ex) -- (1.2ex,0) -- cycle;}
}

\newcommand{\rotatedless}{\rotatebox[origin=c]{270}{$<$}}
\newcommand\vldbdoi{10.14778/3712221.3712238}
\newcommand\vldbpages{729-742}
\newcommand\vldbvolume{18}
\newcommand\vldbissue{3}
\newcommand\vldbyear{2024}
\newcommand\vldbauthors{\authors}
\newcommand\vldbtitle{\shorttitle} 
\newcommand\vldbavailabilityurl{https://github.com/dypoli/computing-hyper-triangle}
\newcommand\vldbpagestyle{empty}

\begin{document}
\title{Efficient Computation of Hyper-triangles on Hypergraphs}

\author{Haozhe Yin}
\affiliation{%
  \institution{The University of New South Wales}
  \streetaddress{Kingsford}
}
\email{unswyhz@gmail.com}

\author{Kai Wang}
\affiliation{%
  \institution{Antai College of Economics and Management, Shanghai Jiao Tong University}
}
\email{w.kai@sjtu.edu.cn}

\author{Wenjie Zhang}
\affiliation{%
  \institution{The University of New South Wales}
  \streetaddress{Kingsford}
}
\email{wenjie.zhang@unsw.edu.au}

\author{Ying Zhang}
\affiliation{%
  \institution{Zhejiang Gongshang University}
}
\email{ying.zhang@zjgsu.edu.cn}

\author{Ruijia Wu}
\affiliation{%
  \institution{Antai College of Economics and Management, Shanghai Jiao Tong University}
}
\email{rjwu@sjtu.edu.cn}

\author{Xuemin Lin}
\affiliation{%
  \institution{Antai College of Economics and Management, Shanghai Jiao Tong University}
}
\email{xuemin.lin@sjtu.edu.cn}

\begin{abstract}
Hypergraphs, which use hyperedges to capture groupwise interactions among different entities, have gained increasing attention recently for their versatility in effectively modeling real-world networks. In this paper, we study the problem of computing hyper-triangles (formed by three fully-connected hyperedges), which is a basic structural unit in hypergraphs. Although existing approaches can be adopted to compute hyper-triangles by exhaustively examining hyperedge combinations, they overlook the structural characteristics distinguishing different hyper-triangle patterns. Consequently, these approaches lack specificity in computing particular hyper-triangle patterns and exhibit low efficiency. In this paper, we unveil a new formation pathway for hyper-triangles, transitioning from hyperedges to hyperwedges before assembling into hyper-triangles, and classify hyper-triangle patterns based on hyperwedges. Leveraging this insight, we introduce a two-step framework to reduce the redundant checking of hyperedge combinations. Under this framework, we propose efficient algorithms for computing a specific pattern of hyper-triangles. Approximate algorithms are also devised to support estimated counting scenarios. Furthermore, we introduce a fine-grained hypergraph clustering coefficient measurement that can reflect diverse properties of hypergraphs based on different hyper-triangle patterns. Extensive experimental evaluations conducted on 11 real-world datasets validate the effectiveness and efficiency of our proposed techniques.

\end{abstract}

\maketitle

\pagestyle{\vldbpagestyle}
\begingroup\small\noindent\raggedright\textbf{PVLDB Reference Format:}\\
\vldbauthors. \vldbtitle. PVLDB, \vldbvolume(\vldbissue): \vldbpages, \vldbyear.\\
\href{https://doi.org/\vldbdoi}{doi:\vldbdoi}
\endgroup
\begingroup
\renewcommand\thefootnote{}\footnote{\noindent
This work is licensed under the Creative Commons BY-NC-ND 4.0 International License. Visit \url{https://creativecommons.org/licenses/by-nc-nd/4.0/} to view a copy of this license. For any use beyond those covered by this license, obtain permission by emailing \href{mailto:info@vldb.org}{info@vldb.org}. Copyright is held by the owner/author(s). Publication rights licensed to the VLDB Endowment. \\
\raggedright Proceedings of the VLDB Endowment, Vol. \vldbvolume, No. \vldbissue\ %
ISSN 2150-8097. \\
\href{https://doi.org/\vldbdoi}{doi:\vldbdoi} \\
}\addtocounter{footnote}{-1}\endgroup

\ifdefempty{\vldbavailabilityurl}{}{
\vspace{.3cm}
\begingroup\small\noindent\raggedright\textbf{PVLDB Artifact Availability:}\\
The source code, data, and/or other artifacts have been made available at \url{\vldbavailabilityurl}.
\endgroup
}

\section{Introduction}

\label{sct:introduction}
\begin{figure}[t]
	\begin{center}
            \subfigure[The original data]{
			\label{fig1.1}
			
			\includegraphics[scale=0.45]{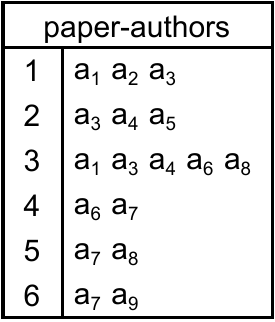}      
		}\hspace{1mm}
		\subfigure[A general graph model]{
			\label{fig1.2}
                
			\includegraphics[scale=0.4]{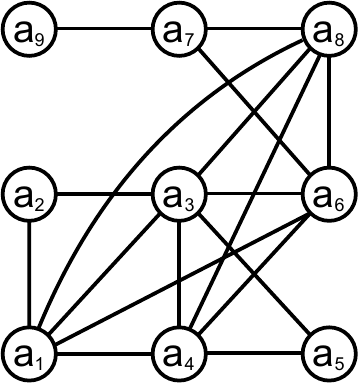}   
		}\hspace{0mm}
            \subfigure[A hypergraph model]{
			\label{fig1.3}
			\includegraphics[scale=0.425]{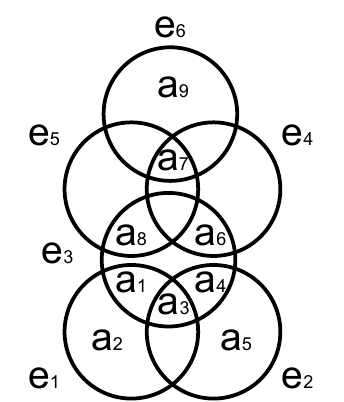}
		}
	\end{center}
        \vspace{-4.5mm}
	\caption{Graph model comparison.}
    
    \label{Graph Comparison}
    \vspace{-6.5mm}
\end{figure}

Hypergraphs are naturally used to model group-based relationships among entities in many real-world applications, such as co-authorship networks \cite{wu2022hypergraph, inoue2022hypergraph, roy2015measuring}, biological networks \cite{feng2021hypergraph, lugo2021classification}, and disease networks \cite{yang2021identifying, ma2022hypergraph}. In technical terms, a hypergraph $G = (V, E)$ is composed of a set of vertices $V$ and a set of hyperedges $E$, where each hyperedge $e \in E$ represents relationships among multiple vertices. Compared with general graph models, hypergraphs can better preserve the integrity of data when describing groupwise intersections. \autoref{Graph Comparison} shows an example to describe the co-authorship relationships using a general graph and a hypergraph. We can intuitively see that the hypergraph depicted in \autoref{Graph Comparison}(c) offers a clear advantage over general graphs by not only displaying the number of papers but also by showcasing the relationships between authors, with each paper serving as the unit of this relational mapping.



\begin{figure}[t]
    \begin{center}
    \centering
    \includegraphics[scale=0.35]{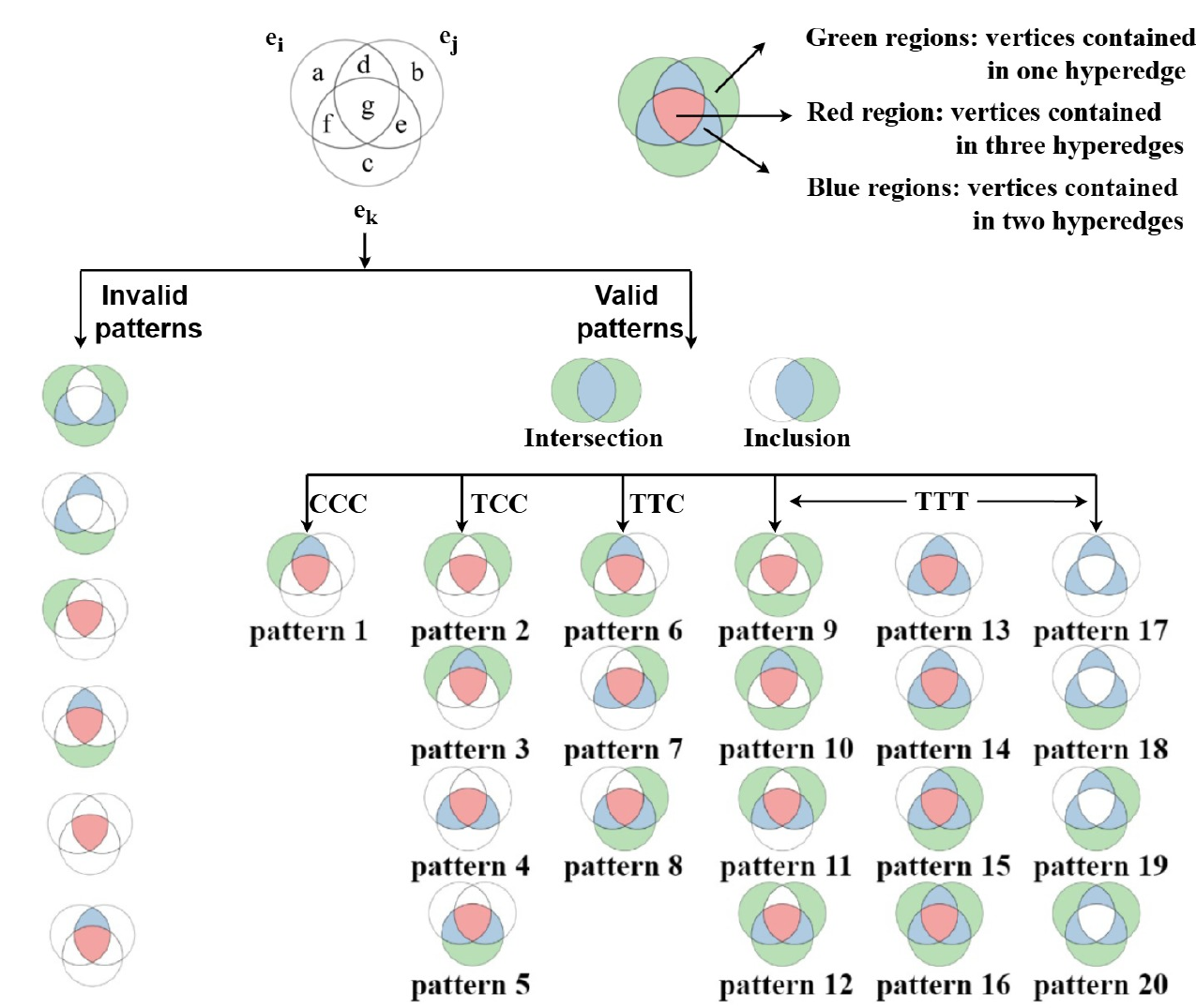} 	
    \end{center}
    \caption{All the patterns for hyper-triangles.}
    \label{Hypergraph Triangle}
    \vspace{-4mm}
\end{figure}

As building blocks of networks/graphs, motifs (i.e., repeated sub-graphs) are essential for graph analysis \cite{wang2020efficient, ma2019linc, gao2022scalable, marcus2010efficient, cai2023efficient, liu2020truss}. Triangles (the smallest non-trivial clique), as a fundamental type of motifs, are typically used to extract information in general graphs \cite{Fang1, Cheng1, Wang1, Yang1}. In hypergraphs, the concept of hyper-triangles (i.e., three pairwise connected hyperedges) has been proposed and proven useful in many applications \cite{nie2021triangle, zhang2023efficiently, gyHori2006triangle}. For instance, in \autoref{Graph Comparison}(c), three hyper-triangles can be identified: \{$e_1$, $e_2$, $e_3$\}, \{$e_3$, $e_4$, $e_5$\}, \{$e_4$, $e_5$, $e_6$\}. These hyper-triangles illustrate complex interaction patterns and higher-order relationships within the network, while traditional triangles merely represent simple pairwise connections among three entities. By considering the relationships among the hyperedges within a hyper-triangle, we can categorize them into various patterns that exhibit distinct internal structures, as depicted in \autoref{Hypergraph Triangle}. In real graphs, the semantic meaning associated with each pattern often varies. For instance, in social networks, different patterns signify distinct social behaviors or organizational structures. In this paper, we focus on the problem of computing hyper-triangles. Specifically, given a hypergraph $G = (V, E)$ and a pattern of hyper-triangles (as shown in \autoref{Hypergraph Triangle}), we aim to find all the hyper-triangles belonging to such a pattern within $G$. Note that by simply combining the algorithms for computing each pattern, we can also identify all the hyper-triangles in $G$. The significance of hyper-triangle computation has been underscored in the literature. Below are some typical examples.




\noindent
\textit{Network measurement.} In hypergraphs, datasets from different domains exhibit distinct local structures due to the properties of data. By computing the frequency of hyper-triangle patterns across different datasets, some patterns within datasets of particular domains can reveal their structural design principles \cite{lee2020hypergraph}. For example, through case studies, we find that in the co-authorship domain, publications from the same research group are more likely to form hyper-triangles of pattern 12 due to hierarchical relationships among authors, while publications from different research groups tend to form hyper-triangles of pattern 10. By analyzing the frequency of these patterns, we can estimate the number of research groups involved in a research field and their collaboration dynamics. Besides, in the email correspondence domain, email accounts within the same organization often exhibit administrative relationships, where a central "organizer" or admin account frequently sends and receives emails to and from all its member accounts. These interactions are more likely to form hyper-triangles of pattern 5, which signify centralized communication. However, when considering external or third-party accounts, due to the randomness of email sending, it is hard to form pattern 5 with these member accounts. By detecting pattern 5 hyper-triangles, we can estimate the number of members within a community. Such an approach can also offer insights for malicious email filtering and anomaly detection. 


\noindent
\textit{Hypergraph clustering coefficient.} Triangle counting is a critical step in computing the clustering coefficient of a graph \cite{seshadhri2014wedge, green2013faster}. Through the clustering coefficient, we can measure the localized closeness and redundancy of a graph. In hypergraphs, the clustering coefficient  \cite{estrada2006subgraph, aksoy2020hypernetwork} is defined as 3$\times |\customtriangle| / |\opentopsquare|$, which requires computing the number of hyper-triangles. Here $\customtriangle$ denotes the set of hyper-triangles and $\opentopsquare$ is the set of open hyper-triangles (i.e., a hyper-triangle in which two hyperedges are disconnected). Note that in hypergraphs, hyper-triangles encompass a variety of patterns as listed in Figure \ref{fig1.2}, which represents different interactions in real scenarios \cite{lee2020hypergraph}. For example, in biological networks \cite{gopalakrishnan2024prediction, hwang2008learning}, a specific pattern reflects a particular reaction chain. Based on this motivation, we further propose a fine-grained clustering coefficient metric (in Section ~\ref{Clustering Coefficient}) to offer users flexibility in adjusting the proportion of various hyper-triangle patterns and reflect different community structures of hypergraphs.

\noindent
\textbf{Motivations and Challenges.} Existing studies \cite{lee2020hypergraph, lee2023temporal} search and categorize all motifs formed by three hyperedges by iteratively enumerating each hyperedge. As a result, when the search target is hyper-triangles of a specific pattern, this traverse-based algorithm still needs to enumerate all hyper-triangles. This process leads to numerous redundant checks of hyperedge combinations, resulting in inefficiencies. In this work, we aim to propose efficient algorithms that can perform targeted computing for specific hyper-triangle patterns, which faces the following challenges.
\vspace{0mm}
\begin{enumerate}[leftmargin=1.8em]
    \item Since each hyperedge can be included in hyper-triangles of various patterns, it is challenging to identify the hyper-triangles of a specific pattern without enumerating hyper-triangles of other patterns.  
    
    \item Existing studies determine the pattern of hyper-triangle by enumerating all the vertices inside, which is very time-consuming. Therefore, it is also a challenge to efficiently identify the pattern of the hyper-triangle.

\end{enumerate}

\noindent
\textbf{Our approaches.} In this paper, we observe that a hyper-triangle can be viewed as a motif composed of three hyperwedges (unordered pairs of connected hyperedges). Therefore, we introduce a new pathway for forming the hyper-triangle, transitioning from two connected hyperedges to a hyperwedge, and then assembling three hyperwedges into a hyper-triangle. Based on the relationships between hyperedges, hyperwedges can be categorized into two types: intersection and inclusion. Building on this classification, we further divide hyper-triangle patterns into four classes depending on the types of three hyperwedges involved. Leveraging this insight, we propose a two-step framework to minimize the redundant computation. For a specific hyper-triangle pattern, the first step is to search and classify all the hyperwedges. Then, based on the types of hyperwedges involved, we search for hyper-triangles within the corresponding hyperwedge set, hence avoiding the enumeration of hyper-triangles of other patterns. 


To address Challenge 2, by pre-saving the common vertices of the two hyperedges contained in each hyperwedge, we can avoid traversing all the vertices in hyper-triangles during the classification stage. Furthermore, by leveraging the structural characteristics of inclusion-type hyperwedges, the patterns involving inclusion-type hyperwedges can be identified in $O(1)$ time. Besides, since most patterns consist of three intersection-type hyperwedges, we further categorize patterns within this class into two subclasses and propose an advanced algorithm to accelerate the search speed for these specific subclasses. Additionally, for the problem of counting hyper-triangles, we also provide an approximation algorithm.

\noindent{\bf Contributions.} In general, our principal contributions are summarized as follows.
\vspace{0mm}
\begin{itemize}[leftmargin=1.2em]

    \item We study the problem of computing hyper-triangles with different patterns to capture the intricate structural characteristics within hypergraphs.

    \item To avoid redundant checking during the searching process, we propose an efficient two-step framework based on a new hyper-triangle formation pathway. We also propose approximate counting algorithms for efficiently estimating the number of hyper-triangles in large hypergraphs.

     \item We propose a fine-grained clustering coefficient model that can reflect diverse properties based on different hyper-triangle patterns. 

    \item We conduct extensive experiments on 11 real hypergraphs. The results demonstrate that the proposed exact algorithms outperform the state-of-the-art algorithm, and the proposed approximation algorithm can achieve more accurate results than the existing algorithm. 
\end{itemize}

\vspace{1mm}

\noindent\textbf{Organization.} The remainder of this paper is structured as follows. Section~\ref{Problem definition} introduces the preliminary and the baseline algorithm. Section~\ref{Exact Algorithm} introduces our proposed exact algorithms as well as the parallel version. All the approximation algorithms are presented in Section~\ref{Approximate Algorithm}. In Section~\ref{Clustering Coefficient}, we propose a fine-grained model of hypergraph clustering coefficient. Extensive experiments are conducted in Section~\ref{Experiments}. In Section~\ref{Related work}, we introduce the related works of this paper. Finally, Section~\ref{Conclusion} concludes this paper.

\section{Problem definition}
\label{Problem definition}

\begin{table}[htb]
\centering
    \resizebox{0.47\textwidth}{!}{%
  \begin{tabular}{|l|l|}
    \hline
    \cellcolor{gray!25}\textbf{Notations} & \cellcolor{gray!25}\textbf{Description}\\
    \hline
    $G = (V, E)$ & \makecell[l] {A hypergraph with vertices $V$ and hyperedges $E$}\\
    \hline
    $E=\{e_1,...,e_{|E|} \}$ & \makecell[l] {The set of hyperedges}\\
    \hline
    $E_v$ & \makecell[l] {The set of all hyperedges containing the vertex $v$}\\
    \hline
    $\rotatedless/ \customtriangle / \opentopsquare$ & \makecell[l] {The set of hyperwedges/hyper-triangles/open hyper-triangles}\\
    \hline
    $ \customtriangle_{p}$ & \makecell[l] {The set of hyper-triangles of pattern $p$}\\
    \hline
    $\rotatedless^t/\rotatedless^c$ & \makecell[l]{The set of all hyperwedges of the intersection/inclusion type}\\
    \hline
    $\rotatedless_{ij}$ & \makecell[l] {A hyperwedge composed of $e_i$ and $e_j$}\\
    \hline
    $\Omega_{ij}$ & \makecell[l] {The set of common vertices between $e_i$ and $e_j$}\\
    \hline
    $\omega_{ij}$ & \makecell[l] {The number of common vertices between $e_i$ and $e_j$}\\
    \hline
    $N_{e_i}$ & \makecell[l] {The set of neighbors of hyperedge $e_i$}\\
    \hline
    $T(\{e_i,e_j,e_k\})$ & \makecell[l] {A hyper-triangle composed of $e_i$, $e_j$ and $e_k$}\\
    \hline
    $H[p]$ & \makecell[l] {The count of the number of hyper-triangles of pattern $p$}\\
    \hline
    \end{tabular}
}
\vspace{2mm}
\caption{Notations}
\label{tab:notation}
\vspace{-6.5mm}
\end{table}

\begin{definition}
[{\bf Hypergraph}] A hypergraph is a graph \( G = (V, E) \) where \( V \) is a set of vertices and $E=\{e_1,...,e_{|E|} \}$ is a set of hyperedges. Each hyperedge $e_i\in E$ is a non-empty set of vertices.
\end{definition}

In this paper, we use $|e_i|$ to denote the number of vertices in the hyperedge $e_i$. Additionally, for any two hyperedges, if they contain the same vertices, we consider them to be connected. We use $N_{e_i}=\{e_{j}\in E:e_i \cap e_j \neq \emptyset\}$ to represent all the hyperedges connected to $e_i$. For vertex $v \in V$, we use $E_v$ to denote all the hyperedges containing the vertex $v$ and we use $E^r_v$ to represent $r\textsuperscript{th}$ hyperedge in $E_v$. Note that, we assume that two different hyperedges do not contain the same set of vertices in this paper.

\begin{definition}
[{\bf Hyperwedge}] Given a hypergraph \( G = (V, E) \) and two hyperedges $e_i$, $e_j \in E$ with $e_i \cap e_j \neq \emptyset$, a hyperwedge $\rotatedless_{ij}$ is a path composed of $e_i$ and $e_j$ in $G$. We use $\Omega_{ij}$ to denote the set of common vertices between $e_i$ and $e_j$ and use $\omega_{ij}$ to denote the number of these vertices.
\end{definition}

For example, in \autoref{Graph Comparison}(c), hyperedges $e_1$ and $e_2$ can form a hyperwedge $\rotatedless_{12}$ since these two hyperedges share a common vertex $a_3$. We denote the set of all hyperwedges as $\rotatedless$. For the hyperedge $e_i$, we use $\rotatedless_{i\_}$ to denote the set for all the hyperwedges containing $e_i$. Besides, we define the order of each hyperwedge $\rotatedless_{ij}$ in $G$ as $\mathcal{O}(\rotatedless_{ij})$ to avoid duplication. For two hyperwedges $\rotatedless_{ij}$ and $\rotatedless_{kl}$, $\mathcal{O}(\rotatedless_{ij})>\mathcal{O}(\rotatedless_{kl})$ if $i>k$ or $i=k$, $j>l$. 





\begin{definition}
[{\bf Intersection/Inclusion hyperwedge}] Given a hypergraph \( G = (V, E) \), for a hyperwedge $\rotatedless_{ij}$ composed of $e_i$ and $e_j$ in $G$, if $e_i \subset e_j$ or $e_j \subset e_i$, then the hyperwedge $\rotatedless_{ij}$ is of the inclusion type; otherwise, it is of the intersection type.
\end{definition}

Similarly, we denote the set of all hyperwedges of the intersection and inclusion types as $\rotatedless^t$ and $\rotatedless^c$ respectively. 

\begin{definition} 
[{\bf Hyper-triangle}] Given a hypergraph \( G = (V, E) \) and three hyperedges $e_i$, $e_j, e_k \in E$ with $e_i \cap e_j \neq \emptyset$, $e_i \cap e_k \neq \emptyset$, and $e_j \cap e_k \neq \emptyset$, a hyper-triangle $T(\{e_i,e_j,e_k\})$ is a subgraph composed of $e_i$, $e_j$ and $e_k$ in $G$.
\end{definition}

For a hyper-triangle $T(\{e_i, e_j, e_k\})$, it encompasses a total of seven regions, which are (a) $ e_i \setminus e_j \setminus e_k $ (b) $ e_j \setminus e_k \setminus e_i $ (c) $ e_k \setminus e_i \setminus e_j $ (d) $ e_i \cap e_j \setminus e_k $ (e) $ e_j \cap e_k \setminus e_i $ (f) $ e_k \cap e_i \setminus e_j $ (g) $ e_i \cap e_j \cap e_k $ as shown in \autoref{Hypergraph Triangle}. Based on the emptiness of each region, after excluding symmetric types, we can divide hyper-triangles into 20 different patterns. Based on the types of the three hyperwedges within each hyper-triangle, we further classify these 20 patterns into four major classes: \textit{CCC}, \textit{TCC}, \textit{TTC} and \textit{TTT} where \textit{T} represents intersection, and \textit{C} represents inclusion. For example, pattern 1 belongs to the \textit{CCC} class because the three hyperwedges forming it are all of the inclusion types. Similarly, patterns 2 to 5 belong to the \textit{TCC} class, patterns 6 to 8 belong to the \textit{TTC} class, and patterns 9 to 20 belong to the \textit{TTT} class. In real datasets, different hyper-triangle patterns can reflect different structural characteristics. Therefore, it is meaningful to search for hyper-triangles of specific patterns.


\noindent
{\bf Problem statement.} Given a hypergraph \( G = (V, E) \) and a given pattern $p$ (i.e., one of the patterns as shown in \autoref{Hypergraph Triangle}), we aim to compute the hyper-triangles of the given pattern $p$ in $G$.

In this paper, we propose exact and approximate algorithms to compute the hyper-triangles. Since the steps of count and enumerate are the same in the exact algorithm, hence the compute means enumerate/count in the exact algorithm. In the approximate algorithm, we estimate the number of hyper-triangles. Note that any hyper-triangle can be captured by one pattern in \autoref{Hypergraph Triangle}, hence we can easily obtain all the hyper-triangles by simply combining algorithms for each pattern.



\vspace{1.5mm}\noindent
{\bf Existing solutions.}
\cite{lee2020hypergraph} introduces an enumeration-based algorithm to find motifs in hypergraphs consisting of three hyperedges, which can also be used to find hyper-triangles.

The details of the algorithm are illustrated in Algorithm \ref{baseline}. The algorithm is based on each hyperedge $e_i$ (Line 3) and checks all its neighboring pairs $\{e_j, e_k\}$ that connect to $e_i$ in a higher order (Line 4), ensuring that each triple $\{e_i, e_j, e_k\}$ is enumerated only once. Note that the algorithm determines whether two hyperedges are connected by pre-constructing a projection graph. If $e_j$, $e_k$ are connected (i.e., $e_j \cap e_k \neq \emptyset$), then it proceeds to identify the pattern of the hyper-triangle $T(\{e_i,e_j,e_k\})$ (Lines 5-7).

\begin{algorithm}[htb]
\small
\caption{Exact-bs} 
\label{baseline} 
    $G(V,E)\leftarrow \text{Input hypergraph}$\\
    $H\leftarrow$Initialize a map to store the number of hyper-triangles\\
    \For{{\bf each} hyperedge $e_i \in E$}{
        \For{{\bf each} unordered hyperedge pair $\{e_j,e_k\} \in \binom{N_{e_i}}{2}$}{
            \If{$e_j \cap e_k \neq \emptyset$ and $i<min\{j,k\}$}{
                $p\leftarrow$Pattern of hyper-triangle $T(\{e_i,e_j,e_k\})$\\
                $H[p]+=1$\\
            }
            
        }
        
    }
    \Return{$H$};
\end{algorithm}

However, Algorithm \ref{baseline} performs redundant checking when searching for hyper-triangles. Specifically, to ensure the correctness of the results, the algorithm must navigate each hyperedge. This process not only consumes considerable time in filtering instances where the three hyperedges only form an open hyper-triangle but also results in the algorithm lacking specificity in computing particular hyper-triangle patterns. Furthermore, constructing the projection graph and identifying the pattern of the hyper-triangle necessitates the repeated enumeration of vertices within each hyperedge, resulting in significant time consumption. For example, consider a sparse hypergraph $G$ with 102 hyperedges and 99 vertices, which contains only three hyper-triangles, as shown in \autoref{A sparse hypergraph}. If the required pattern belongs to the \textit{CCC} class, Algorithm \ref{baseline} needs to go through 104 hyperwedges and filter out 97 open-triangles formed by these hyperwedges to find the hyper-triangle $T(\{e_1,e_2,e_3\})$ of \textit{CCC} class.

    \section{Exact Algorithms}
\label{Exact Algorithm}
\begin{figure}[t]
	\begin{center}
            \includegraphics[scale=0.45]{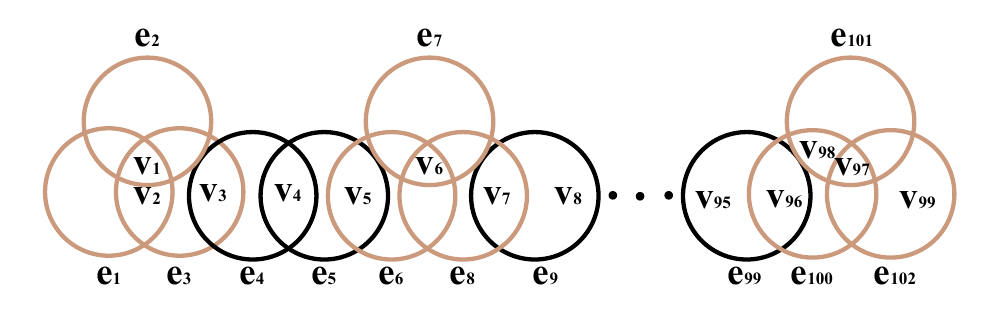}
	\end{center}
        \vspace{-4.5mm}
	\caption{A sparse hypergraph}
    
    \label{A sparse hypergraph}
    \vspace{-6mm}
\end{figure}

To address the issues in the existing algorithm, we introduce a new pathway for constructing hyper-triangles. Rather than forming hyper-triangles directly from hyperedges, this pathway involves transitioning from pairwise connected hyperedges to hyperwedges, and subsequently assembling three hyperwedges into a hyper-triangle. For example, for the hypergraph in \autoref{A sparse hypergraph}, we first identify and classify all the hyperwedges, resulting in 6 inclusion-type hyperwedges and 100 intersection-type hyperwedges. In this case, for hyper-triangles of the \textit{CCC} class, we only need to traverse these 6 inclusion-type hyperwedges to find the hyper-triangle $T(\{e_1,e_2,e_3\})$, which improves the search efficiency compared to Algorithm \ref{baseline}. Building upon this concept, we propose a two-step framework to reduce the redundant checking of hyperedge combinations. Under this framework, we propose efficient exact algorithms for computing hyper-triangle of specific patterns. Additionally, we provide parallel versions of exact algorithms to handle large datasets.


\subsection{The Two-step Framework}

The two-step framework contains the following steps. In the first step, we identify and categorize all hyperwedges into two groups w.r.t. their types (i.e., intersection and inclusion). In the second step, we resort to different algorithms to search for hyper-triangles of specific patterns based on these hyperwedges. 



Algorithm \ref{Preprocess} shows details of the two-step framework. Firstly, for each vertex $v$, we construct a list $E_v$ that saves all hyperedges containing $v$ in ascending order based on the IDs of hyperedges (Line 8). Then, we traverse each hyperedge $e_i$ (Line 9). By using the lists $E_v$ corresponding to all vertices contained in $e_i$, we search for all hyperedges $e_j$ that can form hyperwedges with $e_i$ and record the set of common vertices between $e_i$ and $e_j$ into $\Psi$ (Lines 12-18). Based on the number of vertices contained in $\Psi$, we can determine the type of the hyperwedge $\rotatedless_{ij}$ (Lines 19-22). After enumerating all the hyperedges, all hyperwedges are categorized into two lists based on their types. Finally, based on the given pattern $p$, we employ the corresponding algorithm to find the hyper-triangles. Note that to avoid redundant computations for the same hyperedge in later steps, we store the hyperwedges as follows: for intersection-type hyperwedges $\rotatedless^t_{ij}$, we default to arranging hyperedges with smaller IDs first ($i<j$). For inclusion-type hyperwedges $\rotatedless^c_{ij}$, we default to place the hyperedge with a larger size first ($|e_i|>|e_j|$). 


\begin{algorithm}[htb]

\setstretch{0.9}
\small
\caption{The Two-step Framework} 
\label{Preprocess} 
    $G(V,E)\leftarrow$Hypergraph, $p\leftarrow$pattern of the hyper-triangles\\
    $H\leftarrow$Initialize a map to store the number of hyper-triangles\\
    $\rotatedless^t, \rotatedless^c\leftarrow$Preprocess($G$)  //  step 1\\
    Run the corresponding algorithms for $p$ on the respective hyperwedge lists  //  step 2\\
    {\bf return}\ $H$\\
    
\SetKwFunction{FMain}{Preprocess}
    \SetKwProg{Fn}{Function}{:}{}
    \Fn{\FMain{$G$}}{
        $\rotatedless^c\leftarrow\emptyset$, $\rotatedless^t\leftarrow\emptyset$\\
        $\forall v \in V$ build a list $E_v$ that saves all hyperedges containing $v$\\
        \For{{\bf each} hyperedge $e_i \in$ E}{
            \While{exist hyperedges that can form hyperwedge with $e_i$}{
                $e_j \leftarrow \emptyset$, $n\leftarrow 0$, $\Psi \leftarrow \emptyset$\\
                \For{{\bf each} vertex v $\in e_i$ and $|E_v|>1$}{
                    Remove $e_i$ from $E_v$\\
                    \If{$e_j=\emptyset$ or $e_j>E^1_v$}{
                        $e_j \leftarrow E^1_v$, $n \leftarrow 1$, $\Psi \leftarrow \{v\}$\\
                    }
                    \ElseIf{$e_j=E^1_v$}{
                        $n \leftarrow n+1$, $\Psi \leftarrow \Psi \cup v$\\
                    }
                }
                $\Omega_{ij}\leftarrow\Psi$ //  set of common vertices between $e_i$ and $e_j$\\
                \If{$n<min\{|e_i|, |e_j|\}$}{
                    $\rotatedless^t \leftarrow \rotatedless^t \cup \rotatedless_{ij}$\\
                }
                \ElseIf{$n=min\{|e_i|, |e_j|\}$}{
                    $\rotatedless^c \leftarrow \rotatedless^c \cup \rotatedless_{ij}$\\
                }

            }
        }
        {\bf return}\ $\rotatedless^c, \rotatedless^t$\\
}

\end{algorithm}

\subsection{Algorithms for TTT class}

In this subsection, we present the exact algorithms for searching patterns of the \textit{TTT} class (patterns 9-20). Given that a significant portion of hyper-triangles falls within this class, the efficient computation and classification of these patterns are important. We first propose a basic algorithm for this class. Then, we categorize the \textit{TTT} class into two subclasses and propose more efficient algorithms.


\vspace{1.5mm}\noindent
{\bf{The basic algorithm.}} The details of the basic algorithm are illustrated in Algorithm \ref{Count-TTT}. Initially, we enumerate each intersection-type hyperwedge $\rotatedless_{ij}^t \in \rotatedless^t$, then search for another hyperwedge $\rotatedless_{ik}^t \in \rotatedless^t$ with $\mathcal{O}(\rotatedless^t_{ij})<\mathcal{O}(\rotatedless^t_{ik})$ (Line 3). If they can form a hyper-triangle, we further classify the hyper-triangles (Lines 5-7). Note that in order to determine the pattern of the hyper-triangle $T(\{e_i,e_j,e_k\})$, we need to find the common vertices between $\Omega_{ij}$ and $\Omega_{ik}$. 

\begin{algorithm}
\small
\caption{Count-TTT} 
\label{Count-TTT} 
$G(V,E)\leftarrow$Hypergraph, $\rotatedless^t$, $\rotatedless^c \leftarrow$Preprocess($G$)\\ 
$H\leftarrow$Initialize a map to store the number of hyper-triangles\\
    \For{{\bf each} hyperwedge $\rotatedless^t_{ij} \in\rotatedless^t$ }{
        \For{{\bf each} hyperwedge $\rotatedless^t_{ik} \in\rotatedless^t$ and $\mathcal{O}(\rotatedless^t_{ij})<\mathcal{O}(\rotatedless^t_{ik})$}{ 
            \If{$e_j \cap e_k \neq \emptyset$ and $\omega_{jk}<min\{|e_j|,|e_k|\}$}{
                $p\leftarrow$Pattern of the hyper-triangle $T(\{e_i,e_j,e_k\})$\\
                $H[p]+=1$\\    
            }
        }
        
    }
    {\bf return}\ $H$\\
\end{algorithm}

While Algorithm \ref{Count-TTT} can search hyper-triangles of the \textit{TTT} class, it still faces the following issues. (1) Since the majority of hyper-triangles belong to the \textit{TTT} class, which includes 12 patterns, the runtime of the algorithm is still relatively long if we wish to find hyper-triangles of a specific pattern in this class. (2) The algorithm spends a significant amount of time filtering out the hyperedges that can only form an open hyper-triangle. To further enhance the efficiency of the algorithm, it is important to avoid traversing the open hyper-triangles. (3) The above process computes hyper-triangles by iteratively enumerating each hyperwedge. As the dataset grows, this method becomes increasingly time-consuming. Therefore, avoiding iteratively checking each hyperwedge would further enhance the efficiency of the algorithm.

To tackle the issue (1), the patterns within the \textit{TTT} class are further divided into two subclasses. If there exists a set of vertices contained by all three hyperedges in a hyper-triangle pattern, then such a pattern is categorized into the \textit{DenseTTT} subclass (patterns 9-16). Otherwise, they are classified into the \textit{SparseTTT} subclass (patterns 17-20).

\vspace{1.5mm}\noindent
{\bf{The algorithm for DenseTTT subclass.}} To address the issue (2), we observe that the three hyperedges contained in an open hyper-triangle do not share any common vertices, which is opposite to the structure of the hyper-triangles of the \textit{DenseTTT} subclass. This implies that for an intersection-type hyperwedge $\rotatedless^t_{ij}$, if we can find other intersection-type hyperwedges $\rotatedless^t_{ik}$ such that $\Omega_{ij} \cap \Omega_{ik} \neq \emptyset$, then they can definitely form hyper-triangles, hence we can avoid traversing open hyper-triangles. To achieve this method, for a hypergraph $G=(E, V)$, we initially construct a list $\tau_v$ for each vertex $v\in V$ which maps $v$ to all the intersection-type hyperwedges $\rotatedless_{ij}^t$ where $v \in \Omega_{ij}$. Then, for each intersection-type hyperwedge $\rotatedless^t_{ij} \in \rotatedless^t$, based on the list $\tau_v$ of each vertex $v \in \Omega_{ij}$, we search for other intersection-type hyperwedges $\rotatedless^t_{ik}$ where $\mathcal{O}(\rotatedless^t_{ij})<\mathcal{O}(\rotatedless^t_{ik})$ (we denote the set of these hyperwedges for $\rotatedless^t_{ij}$ as $S_{ij}$). Since the hyperwedges in $S_{ij}$ contain the same hyperedge $e_i$ with $\rotatedless^t_{ij}$, all of them can form a hyper-triangle with $\rotatedless^t_{ij}$.

\begin{lemma}
For two hyperwedges $\rotatedless^t_{ij}, \rotatedless^t_{ik}\in \rotatedless^t$ where $\mathcal{O}(\rotatedless^t_{ij})<\mathcal{O}(\rotatedless^t_{ik})$. if the set of common vertices between $e_i$ and $e_j$ equals to the set of common vertices between $e_i$ and $e_k$ (i.e., $\Omega_{ij}=\Omega_{ik}$), then $S_{ij} \supset S_{ik}$.
\label{Skip}
\end{lemma}\vspace{-0.5mm}

\noindent
{\bf Proof:} For each hyperwedge $\rotatedless^t_{il}\in S_{ik}$, we have $\mathcal{O}(\rotatedless^t_{ij})<\mathcal{O}(\rotatedless^t_{ik})<\mathcal{O}(\rotatedless^t_{il})$. Since $\rotatedless^t_{il}$ is found through the vertices $v \in \Omega_{ik}$, and $\Omega_{ij}=\Omega_{ik}$, hence $\rotatedless^t_{il}$ will also be found when searching for hyperwedges through the vertex $v \in \Omega_{ij}$. Therefore, $S_{ij} \supset S_{ik}$.
\vspace{1.5mm}


\begin{algorithm}[htb]
\setstretch{0.9}
\small
\caption{Count-DenseTTT} 
\label{Count-DenseTTT} 
$G(V,E)\leftarrow$Hypergraph, $\rotatedless^t$, $\rotatedless^c \leftarrow$Preprocess($G$)\\
$H\leftarrow$Initialize a map to store the number of hyper-triangles\\
    $\forall v \in V$ Initialize a list $\tau_v$\\
    \For{{\bf each} hyperwedge $\rotatedless^t_{ij} \in\rotatedless^t$ }{
        \For{{\bf each} vertex $v \in \Omega_{ij}$ }{ 
            $\tau_{v} \leftarrow \tau_{v} \cup \rotatedless^t_{ij}$\\
        } 
    }
    \For{{\bf each} hyperwedge $\rotatedless^t_{ij} \in\rotatedless^t$ and $\rotatedless^t_{ij}$ has not been visited}{ 
        $\Phi\leftarrow \emptyset$, $\Phi \leftarrow \Phi \cup \rotatedless^t_{ij}$\\
        \While{exist hyperwedges that form hyper-triangles with $\rotatedless^t_{ij}$}{
            $w_1\leftarrow \emptyset$, $\alpha\leftarrow 0$\\
            \For{{\bf each} vertex $v \in \Omega_{ij}$ and $|\tau_{v}|>1$ }{ 
                remove $\rotatedless^t_{ij}$ from $\tau_{v}$\\
                $\rotatedless^t_{kl}\leftarrow$a hyperwedge in $\tau_{v}$ with smallest order and contain same hyperedge with $\rotatedless^t_{ij}$ (i.e., $k=i$)\\
                \If{$w_1 = \emptyset$ or $\mathcal{O}(w_1)>\mathcal{O}(\rotatedless^t_{kl})$}{
                    $w_1 \leftarrow \rotatedless^t_{kl}$, $\alpha \leftarrow 1$\\
               }
               \ElseIf{ $\mathcal{O}(w_1)=\mathcal{O}(\rotatedless^t_{kl})$}{
                    $\alpha+=1$\\
               }

            } 

            \For{{\bf each} hyperwedge $w_2  \in \Phi$ }{ 
                 $p\leftarrow$pattern of hyper-triangle formed by $w_1$ and $w_2$\\
                 $H[p]+=1$\\
                
            }
            \If{$\alpha=\omega_{ij}$}{
                $\Phi \leftarrow \Phi \cup w_1$ and mark $w_1$ as visited\\
            }
            
        }    
    }
    {\bf return}\ $H$\\
    
\end{algorithm}

We can address the issue (3) based on Lemma \autoref{Skip}. Specifically, during the aforementioned traversal process, we incorporate the following step: for the current hyperwedge $\rotatedless^t_{ij}$, whenever another hyperwedge $\rotatedless^t_{ik}$ is found so that $\Omega_{ij}=\Omega_{ik}$, we save $\rotatedless^t_{ik}$. After that, every hyperwedge found that can form a hyper-triangle with $\rotatedless^t_{ij}$ will also be able to form a hyper-triangle with $\rotatedless^t_{ik}$.

Based on the above optimization, we obtain Algorithm \ref{Count-DenseTTT}. Firstly, we build the list $\tau_v$ for each vertex $v$ (Lines 4-6). Then, we iteratively enumerate each hyperwedge. For the current hyperwedge $\rotatedless^t_{ij}$, if it has been visited, we skip it (Line 7). Otherwise, based on the lists $\tau_v$ for each vertex $v \in \Omega_{ij}$, we search for another hyperwedges $w_1$ (Lines 11-17). If $w_1$ can form a hyper-triangle with $\rotatedless^t_{ij}$, then it can also form hyper-triangles with the hyperwedges stored in $\Phi$ (lines 18-20). Finally, if the set of common vertices between the two hyperedges in hyperwedge $w_1$ equal to the set $\Omega_{ij}$, then we add $w_1$ to $\Phi$ and mark it as visited (Lines 21-22).

\vspace{1.5mm}\noindent
{\bf{The algorithm for SparseTTT subclass.}} The \textit{SparseTTT} subclass includes four patterns (patterns 17-20). In this subclass, the three hyperedges within each pattern do not share any common vertices, which means we cannot use vertices to identify hyperwedges. Therefore, the only way to search for hyper-triangles of this subclass is through a similar approach in \textit{Count-TTT}. By checking whether three hyperedges contain the same vertices, we can rapidly filter out hyper-triangles of the \textit{Dense-TTT} subclass, thereby speeding up the search process. However, the time savings from this approach are minimal, making the search for hyper-triangles of the \textit{SparseTTT} subclass the most time-consuming task.

\subsection{Algorithms for other classes}

In this subsection, we introduce algorithms for the remaining three classes: \textit{CCC} (pattern 1), \textit{TCC} (patterns 2-5), and \textit{TTC} (patterns 6-8). Unlike the \textit{TTT} class, these three classes each contain at least one inclusion-type hyperwedge. Therefore, by leveraging the structural characteristics of inclusion-type hyperwedge, we can determine the pattern of the hyper-triangle in $O(1)$ time. This approach avoids the process of determining the pattern of the hyper-triangle by finding common vertices among the three hyperedges in it.

\vspace{1.5mm}\noindent
{\bf{The algorithm for CCC class.}} We begin with the \textit{CCC} class (pattern 1), which consists of one pattern formed by three inclusion-type hyperwedges. The details are presented in lines 4-8 of Algorithm \ref{Count-rest}. Firstly, we traverse the hyperwedges in $\rotatedless^c$. For the current hyperwedge $\rotatedless^c_{ij}$, we search for another hyperwedge $\rotatedless^c_{ik}$. If $e_j \supset e_k$, then there is a hyper-triangle $T(\{e_i,e_j,e_k\})$ of pattern 1 (denoted as $p_1$ in Algorithm \ref{Count-rest}).

\vspace{1.5mm}\noindent
{\bf{The algorithm for TCC class.}} The \textit{TCC} class includes four patterns (patterns 2-5). By analyzing the structure of the patterns within the \textit{TCC} class, we can efficiently find and classify the hyper-triangles that belong to this class. Specifically, patterns 2 and 3 are constructed from two intersecting hyperedges, accompanied by a third hyperedge that is inclusively contained by the first two. Therefore, to find hyper-triangles of patterns 2 and 3, we first enumerate all intersection-type hyperwedges $\rotatedless^t_{ij}$ (Line 10). Then, for each hyperwedge $\rotatedless^t_{ij}$, we search for hyperedges $e_k$ that are contained by both $e_i$ and $e_j$ (Lines 11-12). For the found hyper-triangle $T(\{e_i,e_j,e_k\})$, if the number of common vertices between $e_i$ and $e_j$ is greater than the number of vertices in $e_k$ (i.e., $\omega_{ij}>|e_k|$), then it belongs to pattern 3 (Lines 13-14). Otherwise, it belongs to pattern 2 (Lines 15-16). For patterns 4 and 5, the structure consists of two intersecting hyperedges, complemented by an additional hyperedge that includes both intersecting ones within it. Similarly, we enumerate all intersection-type hyperwedges $\rotatedless^t_{ij}$ (Line 10). Then, for each hyperwedge $\rotatedless^t_{ij}$, we search for hyperedges $e_k$ that contain both $e_i$ and $e_j$ (Lines 17-18). For the found hyper-triangle $T(\{e_i,e_j,e_k\})$, if $\omega_{ij}+\omega_{ik}-\omega_{jk}=|e_i|$, then it belongs to pattern 4 (Lines 19-20). Otherwise, it belongs to pattern 5 (Lines 21-22).

\begin{algorithm}[htb]
\setstretch{0.82}
\small
\caption{Count-TTC/TCC/CCC} 
\label{Count-rest} 
$G(V,E)\leftarrow$Hypergraph, $\rotatedless^t$, $\rotatedless^c \leftarrow$Preprocess($G$)\\
$p\leftarrow$pattern of the hyper-triangles\\
$H\leftarrow$Initialize a map to store the number of hyper-triangles\\
        \If{$p \in \text{CCC class}$}{
            \For{{\bf each} hyperwedge $\rotatedless^c_{ij} \in\rotatedless^c$ }{
                \For{{\bf each} hyperwedge $\rotatedless^c_{ik} \in\rotatedless^c$ }{
                    \If{$e_j \supset e_k$}{
                        $H[p_1]+=1$\\
                    }
                }
            }
        }
        \If{$p \in \text{TCC class}$}{
            \For{{\bf each} hyperwedge $\rotatedless^t_{ij} \in\rotatedless^t$ }{
                \For{{\bf each} hyperwedge $\rotatedless^c_{ik} \in\rotatedless^c$ }{ 
                    \If{$e_j \supset e_k$}{
                        \If{$\omega_{ij}>|e_k| $ }{
                            $H[p_3]+=1$\\
                        }
                        \Else{
                            $H[p_2]+=1$\\
                        }
                    }
                }
                \For{{\bf each} hyperwedge $\rotatedless^c_{ki} \in\rotatedless^c$ }{ 
                    \If{$e_k \supset e_j$}{
                        \If{$\omega_{ki}+\omega_{kj}-\omega_{ij}=|e_k| $ }{
                            $H[p_4]+=1$\\
                        }
                        \Else{
                            $H[p_5]+=1$\\
                        }
                    }
                }  
            }
        } 
        \If{$p \in \text{TTC class}$}{
            \For{{\bf each} hyperwedge $\rotatedless^c_{ij} \in\rotatedless^c$ }{
                \For{{\bf each} hyperwedge $\rotatedless^t_{ik}, \rotatedless^t_{ki} \in\rotatedless^t$ }{ 
                    \If{$e_j \cap e_k \neq \emptyset$ and $\omega_{jk}<min\{|e_j|,|e_k|\}$}{
                        \If{$\omega_{ik}=\omega_{jk} $}{
                            $H[p_6]+=1$\\
                        }
                        \Else{
                            \If{$\omega_{ij}+\omega_{ik}-\omega_{jk}=|e_i|$ }{
                                $H[p_7]+=1$\\
                            }
                            \Else{
                                $H[p_8]+=1$\\
                            }
                        }
                    }
                }
                
            }
        }
    {\bf return}\ $H$\\
\end{algorithm}

\vspace{1.5mm}\noindent
{\bf{The algorithm for TTC class.}} The \textit{TTC} class includes three patterns (patterns 6-8), and the search process is similar to the \textit{TCC} class. As illustrated in lines 23-33 of Algorithm \ref{Count-rest}. First, we enumerate each inclusion-type hyperwedge $\rotatedless^c_{ij} \in \rotatedless^c$ and search for hyperedges $e_k$ that intersect with both $e_i$ and $e_j$ (Lines 24-26). For the found hyper-triangle $T({e_i,e_j,e_k})$, if the number of common vertices between $e_i$ and $e_k$ equals the number of common vertices between $e_j$ and $e_k$ (i.e., $\omega_{ik} = \omega_{jk}$), then this hyper-triangle belongs to pattern 6 (Lines 27-28). If not, we compute the value of $\omega_{ij} + \omega_{ik} - \omega_{jk}$. If it equals the number of vertices in $e_i$, then the hyper-triangle belongs to pattern 7 (Lines 30-31); otherwise, it belongs to pattern 8 (Lines 32-33).


\subsection{Complexity Analysis}

\begin{table}[htb]
    
    \centering
    \resizebox{0.47\textwidth}{!}{%
    \begin{tabular}{|l|l|l|}
        \hline
        \multicolumn{1}{|c|}{\cellcolor{gray!25}\textbf{Algorithms}} & 
        \multicolumn{1}{c|}{\cellcolor{gray!25}\textbf{Preprocessing} $(T_{1})$} & 
        \multicolumn{1}{c|}{\cellcolor{gray!25}\textbf{Searching and Classifying} $(T_{2})$} \\
        \hline
        \textit{Exact-bs} & $O(\sum_{e_{i} \in E}(|e_i|\cdot|N_{e_i}|))$ & $O(\sum_{e_i\in E}(|N_{e_i}|^2\cdot |e_i|))$ \\
        \hline 
        \textit{Count-TTC} & $O(\sum_{\rotatedless_{ij} \in \rotatedless}(|e_i|+|e_j|))$ & $O(\sum_{\rotatedless_{ij}^c \in \rotatedless^c}(|\rotatedless_{i\_}^t|+|\rotatedless_{\_i}^t|))$\\
        \hline
        \textit{Count-TCC} & $O(\sum_{\rotatedless_{ij} \in \rotatedless}(|e_i|+|e_j|))$ & $O(\sum_{\rotatedless_{ij}^t \in \rotatedless^t}(|\rotatedless_{i\_}^c|+|\rotatedless_{\_i}^c|))$ \\
        \hline
        \textit{Count-CCC} & $O(\sum_{\rotatedless_{ij} \in \rotatedless}(|e_i|+|e_j|))$ & $O(\sum_{\rotatedless_{ij}^c \in \rotatedless^c}(|\rotatedless_{i\_}^c|))$ \\
        \hline
        \textit{Count-TTT} & $O(\sum_{\rotatedless_{ij} \in \rotatedless}(|e_i|+|e_j|))$ & $O(\sum_{\rotatedless_{ij}^t \in \rotatedless^t}(|\rotatedless_{i\_}^t|\cdot \omega_{ij}))$ \\
        \hline 
         \textit{Count-DenseTTT} & $O(\sum_{\rotatedless_{ij} \in \rotatedless}(|e_i|+|e_j|))$ & $O(\sum_{\rotatedless_{ij}^t \in \rotatedless^t}(|S_{ij}|\cdot \omega_{ij}))$ \\
        \hline
        \textit{Count-SparseTTT} & $O(\sum_{\rotatedless_{ij} \in \rotatedless}(|e_i|+|e_j|))$ & $O(\sum_{\rotatedless_{ij}^t \in \rotatedless^t}(|\rotatedless_{i\_}^t|\cdot \omega_{ij}))$\\
        \hline
    \end{tabular}%
    }

    \begin{flushleft}
    \footnotesize
    {* The total time complexity = $T_1$+$T_2$}
    
    \end{flushleft}

    \vspace{-0mm}
    \caption{Time complexity}
    \label{Time complexity} 
    \vspace{-4mm}
\end{table}

\autoref{Time complexity} shows the time complexity of all exact algorithms in the two stages. First, in the preprocessing stage, \textit{Exact-bs} examines each hyperedge $e_i$ and enumerates the vertices within $e_i$ to identify all its neighbors $N_{e_i}$. As a result, the time complexity of this process is $O(\sum_{e_{i} \in E}(|e_i|\cdot|N_{e_i}|))$. However, the proposed algorithms not only identify all hyperwedges $\rotatedless_{ij}$ but also determine the common vertices between the two hyperedges in $\rotatedless_{ij}$. Therefore, the time complexity of the proposed approach is $O(\sum_{\rotatedless_{ij} \in \rotatedless}(|e_i| + |e_j|))$. 

After preprocessing, \textit{Exact-bs} examines every pair of neighbors for each hyperedge $e_i$. As a result, it traverses $O(\sum_{e_i \in E} |N_{e_i}|^2)$ triples. For each hyper-triangle, \textit{Exact-bs} requires $O(|e_i|)$ time to determine its pattern. Therefore, the time complexity at this stage is $O(\sum_{e_i \in E}(|N_{e_i}|^2 \cdot |e_i|))$. For the proposed algorithms, they perform the search based on hyperwedges. For example, \textit{Count-TTC} first traverses each inclusion-type hyperwedge $\rotatedless_{ij}^c$, and for each $\rotatedless_{ij}^c$, it searches for all other intersection-type hyperwedges that contain the hyperedge $e_i$. Therefore, it traverses $O(\sum_{\rotatedless_{ij}^c \in \rotatedless^c} (|\rotatedless_{i\_}^t| + |\rotatedless_{\_i}^t|))$ hyperwedge pairs (note that $\rotatedless_{i\_}^t$ is the set of all hyperwedges $\rotatedless_{ik}$ where $i < k$). For each hyper-triangle, \textit{Count-TTC} only requires $O(1)$ time to determine its pattern. hence the time complexity is $O(\sum_{\rotatedless_{ij}^c \in \rotatedless^c}(|\rotatedless_{i\_}^t|+|\rotatedless_{\_i}^t|))$. For both \textit{Count-TCC} and \textit{Count-CCC}, they can also determine the hyper-triangle pattern in $O(1)$ time, therefore their time complexity is similar to that of \textit{Count-TTC}. For \textit{Count-TTT}, \textit{Count-DenseTTT}, and \textit{Count-SparseTTT}, for each hyperwedge pair $\{\rotatedless_{ij}, \rotatedless_{ik}\}$, they need to enumerate the vertices in $\Omega_{ij}$ to determine the pattern of the hyper-triangle, hence their time complexities are $O(\sum_{\rotatedless_{ij}^t \in \rotatedless^t}(|\rotatedless_{i\_}^t|\cdot \omega_{ij}))$, $O(\sum_{\rotatedless_{ij}^t \in \rotatedless^t}(|S_{ij}|\cdot \omega_{ij}))$, $O(\sum_{\rotatedless_{ij}^t \in \rotatedless^t}(|\rotatedless_{i\_}^t|\cdot \omega_{ij}))$ respectively.


For the space complexity of \textit{Exact-bs}, it requires $O(\sum_{e_i\in E}(|e_i|) +|\rotatedless|)$ space to store all the hyperedges and the projection graph. As for the proposed algorithms, in addition to storing all the hyperedges, they also need to store the hyperwedges and the common vertices between the two hyperedges in each hyperwedge. Therefore, the space complexity is $O(\sum_{e_i\in E}(|e_i|)+\sum_{\rotatedless_{ij} \in \rotatedless}(\omega_{ij}))$.

\subsection{Parallelization}
Note that our algorithms are also easily to be parallelized. Existing studies \cite{lee2020hypergraph} convert the exact algorithm into a parallel form by parallelizing hypergraph projection. This allows multiple threads to independently process different hyperedges concurrently (denoted as \textit{Par-bs}). In our algorithm, we initially use multiple threads to concurrently process different hyperedges and extract all hyperwedges along with their types and the common vertices between the two hyperedges involved. Once all threads stop, we aggregate the information for hyperwedges obtained from each thread together. Then, for each exact algorithm, we use multiple threads to process the respective hyperwedges in parallel (denoted as \textit{Par-adv}).

\section{Approximate Counting}
\label{Approximate Algorithm} 
In this section, we propose approximate algorithms that can efficiently estimate the number of hyper-triangles of various patterns in hypergraphs. Currently, commonly used estimation methods resort to random sampling approaches for motifs \cite{zhang2023scalable, seshadhri2013triadic, rahman2014graft, jha2015path, pavan2013counting, lim2015mascot}, and there are three types of sampled elements including vertex sampling, edge sampling, and wedge sampling. Similarly, in hypergraphs, we can employ vertex sampling, hyperedge sampling, and hyperwedge sampling for approximation. In general, the workflow of hypergraph sampling algorithm is as follows: (1) Extract a subgraph from the entire graph. (2) Sample some hyperedges, hyperwedges, and vertices from the subgraph, and for each hyperedge or hyperwedge, count the number of hyper-triangles where its ID/order is the smallest; for each vertex, identify all hyperedges containing it, and count the number of hyper-triangles where the IDs of those hyperedges are the smallest. (3) Based on the number of sample elements and the size of the subgraph, compute the number of hyper-triangles in the original graph. However, although these three sampling methods can be used to estimate the number of hyper-triangles, their results still exhibit differences. In fact, since each hyperedge contains a different number of vertices, the results obtained from vertex sampling are biased. The results from hyperedge and hyperwedge sampling, while unbiased, also have different variances. Hence, we present the corresponding proof and propose an approximation algorithm based on the sampling method with the smallest variance.


\begin{lemma}
\label{Vsampling}
Using vertex sampling to estimate the number of hyper-triangles is not unbiased, while using hyperedge sampling or hyperwedge sampling to estimate the number of hyper-triangles is unbiased.
\end{lemma}

\noindent
{\bf Proof: } We use the element to refer to a vertex, hyperedge, or hyperwedge, and we use $\delta$ to refer to the total number of the elements ($|V|$ for vertex, $|E|$ for hyperedge, $|\rotatedless|$ for hyperwedge). We denote the hyperedge with the lowest ID in the $j^{th}$ hyper-triangle as $\Gamma_j$. Besides, we use $X^p_{ij}=1$ to represent that $j^{th}$ hyper-triangle of pattern $p$ is being counted for the $i^{th}$ element, otherwise $X^p_{ij}=0$. For the $i^{th}$ vertex, the probability of it being included in the $j^{th}$ hyper-triangle is $\frac{|\Gamma_j|}{|V|}$. So, if we sample $\alpha$ vertices, by the linearity of expectation, its expected value is: $\mathrm{E}[X]=\sum_{i=1}^{\alpha}\sum_{j=1}^{H[p]} \mathrm{E}[X^p_{ij}]=\frac{\alpha}{|V|} \sum_{j=1}^{H[p]}|\Gamma_j|$. Finally, based on the proportion of $\alpha$ to $|V|$, the result is obtained as: $\frac{\alpha}{|V|}\sum_{j=1}^{H[p]}|\Gamma_j|\times \frac{|V|}{\alpha}=\sum_{j=1}^{H[p]}|\Gamma_j|\neq H[p]$. Therefore, the results obtained through this method are not unbiased. However, since the probability of $i$-th hyperedge or hyperwedge being included in the $j^{th}$ hyper-triangle is $\frac{1}{\delta}$, if we sample $\alpha$ hyperedges or hyperwedges, its expected value is $\mathrm{E}[X]=\sum_{i=1}^{\alpha}\sum_{j=1}^{H[p]} \frac{1}{\delta}=\frac{\alpha H[p] }{\delta}$. Based on the proportion of $\alpha$ to $\delta$, we can obtain an unbiased result: $\frac{\alpha H[p] }{\delta}\times \frac{\delta}{\alpha}=H[p]$. Therefore, the results obtained through these two methods are unbiased.

\vspace{1.5mm}
Due to the property of hypergraph, the results obtained through vertex sampling are not unbiased. Therefore, we proceed to compare the variance between hyperedge sampling and hyperwedge sampling.

\begin{lemma}
\label{variance}
The variance of the results obtained by hyperwedge sampling is less than that obtained by hyperedge sampling.
\end{lemma}

\noindent
{\bf Proof: } Similarly, we use the element to refer to a hyperedge, or hyperwedge. The variance that hyper-triangles of pattern $p$ are counted while processing a sampled element can be expressed as $\mathrm{Var}[\sum_{j=1}^{H[p]}X^p_{ij}]$. Besides, since the sampling is done uniformly at random, hence $\mathrm{Cov}[X^p_{ij},X^p_{kl}]=0$. Therefore, when we sample $\alpha$ elements, their variance can be expressed and decomposed as line 1 of \autoref{var}. Next, since $(X^p_{ij})^2=X^p_{ij}$ and through the relationships between variance, covariance, and expectation, we can further decompose the expression as: 

\vspace{-2mm}

\small
\begin{equation}
\footnotesize
\label{var}
\begin{aligned}
    &\mathrm{Var}\left[\sum_{i=1}^{\alpha}\sum_{j=1}^{H[p]}X^p_{ij}\right] = \sum_{i=1}^{\alpha}\left(\sum_{j=1}^{H[p]}\mathrm{Var}[X^p_{ij}]+\sum_{j\neq k}\mathrm{Cov}(X^p_{ij}, X^p_{ik})\right)\\
    &=\sum_{i=1}^{\alpha}\left(\sum_{j=1}^{H[p]}\left(\mathrm{E}[(X^p_{ij})^2]-\mathrm{E}[X^p_{ij}]^2\right)+\sum_{j\neq k}\left(\mathrm{E}[X^p_{ij}\cdot X^p_{ik}]-\mathrm{E}[X^p_{ij}]\mathrm{E}[X^p_{ik}]\right)\right)\\
    &=\alpha\left(H[p]\left(\frac{1}{\delta}-\frac{1}{\delta^2}\right)+\sum_{j\neq k}\mathrm{E}[X^p_{ij}\cdot X^p_{ik}]-\sum_{j\neq k}\frac{1}{\delta^2}\right)
\end{aligned}
\end{equation}
\normalsize

\noindent
We can further decompose the expected value $\mathrm{E}[X^p_{ij} \cdot X^p_{ik}]$ as follow: $\mathrm{E}[X^p_{ij} \cdot X^p_{ik}]=\mathrm{P}[X^p_{ij}=1]\mathrm{P}[X^p_{ik}=1|X^p_{ij}=1]$. For the $j^{th}$ and $k^{th}$ hyper-triangle, if they do not contain any common element, it is impossible to sample them at the same time. In this case, $\mathrm{E}[X^p_{ij}\cdot X^p_{ik}] = 0$. Since we only count the hyper-triangle in which the ID/order of the hyperedge/hyperwedge is the smallest. Therefore, for two hyper-triangles that contain the same elements, the probability of being sampled simultaneously is $\frac{1}{3}$. In this case, $\mathrm{E}[X^p_{ij}\cdot X^p_{ik}] = \frac{1}{\delta}\cdot\frac{1}{3}$. The variance can be expressed as: $\mathrm{Var}\left[\sum_{i=1}^{\alpha}\sum_{j=1}^{H[p]}X^p_{ij}\right]=\alpha\left(H[p]\left(\frac{1}{\delta}-\frac{1}{\delta^2}\right)+\frac{\gamma_{p}'}{3\delta}-\frac{\gamma_{p}}{\delta^2}\right)$, where $\gamma_p$ is total pairs of hyper-triangle of pattern $p$, and $\gamma_p'$ is total pairs of hyper-triangle of pattern $p$ that share the same hyperedge/hyperwedge.

\vspace{1.5mm}\noindent
{\bf Comparing the two sampling methods.} When we sample the same proportion of hyperedges/hyperwedges (i.e., $\frac{\alpha}{\delta}=c$ where $c$ is a constant), the variance can be simplified as $H[p](c-\frac{c}{\delta})+\frac{c\gamma_{p}'}{3}-\frac{c\gamma_{p}}{\delta}$. In real datasets, the term $\frac{c\gamma_{p}'}{3}$ is significantly larger than the other two terms. Therefore, we only need to compare the magnitudes of this term. It is obvious that the number of hyper-triangle pairs sharing the same hyperedge is greater. Hence, the results obtained by hyperwedge sampling are more accurate.

\subsection{The Basic Algorithm}

Through the above comparison, we conclude that hyperwedge sampling is unbiased and has smaller errors, making it the most suitable approximation method. In \cite{lee2020hypergraph}, researchers use hyperwedge sampling to estimate the number of motifs in a hypergraph. However, since their algorithm involves estimating the number of open hyper-triangles and the estimation process includes redundant checking of vertices. Therefore, we propose an algorithm specifically designed for estimating hyper-triangles. Additionally, we incorporate Preprocess to reduce redundant checking. To accommodate datasets of various sizes, we introduce a parameter $\sigma$, where $\sigma \in (0, 1]$. Depending on the value of this parameter, we extract subgraphs from the original graph for estimation. This approach helps avoid excessive computation times in large datasets.


The details of the approximate algorithm are presented in Algorithm \ref{Appro-bs}. Firstly, we uniformly at random select $\sigma|E|$ hyperedges from the original graph to obtain the subgraph $G'$ (Line 2). Then, we sample $\alpha$ hyperwedges from the subgraph (Line 4). For each hyperwedge $\rotatedless_{ij}$, we search for all hyperwedges $\rotatedless_{ik}$ where $\mathcal{O}(\rotatedless^t_{ij})<\mathcal{O}(\rotatedless^t_{ik})$ that can form hyper-triangles with it (Lines 7-9). Finally, we determine the number of hyper-triangles in the subgraph based on the ratio of $\alpha$ to $|\rotatedless|$, and then scale it up to estimate the number of hyper-triangles in the original graph based on the ratio of $|E'|^3$ to $|E|^3$ (Lines 10-11).

\begin{algorithm}[htb]

\small
\caption{Appro-bs} 
\label{Appro-bs} 
    $G(V,E)\leftarrow$Hypergraph, $\sigma \leftarrow$sampling proportion, $\alpha \leftarrow$number of samples, $\hat{H}\leftarrow$map to store the estimate count of each pattern\\
    $G'(V',E')\leftarrow$sample $\sigma|E|$ hyperedges from $G$ uniformly at random\\
     $\rotatedless^t$, $\rotatedless^c \leftarrow$Preprocess($G'$)\\
     \For{$n\leftarrow 1 \dots \alpha$}{
        $\rotatedless_{ij} \leftarrow$sample a hyperwedge from $G'$ uniformly at random\\
        \For{{\bf each} hyperwedge $\rotatedless_{ik} \in\rotatedless$ }{
            \If{$\mathcal{O}(\rotatedless^t_{ij})<\mathcal{O}(\rotatedless^t_{ik})$ and $e_j \cap e_k \neq \emptyset$}{
                $p\leftarrow$Pattern of hyper-triangle $T(\{e_i,e_j,e_k\})$\\
                $\hat{H}[p]+=1$
           }
        }
     }
     \For{$p\leftarrow 1 \dots 20$}{
        $\hat{H}[p]=\hat{H}[p] \cdot \frac{|\rotatedless|}{\alpha  }\cdot \sigma^{-3}$
     }
    {\bf return}\ $\hat{H}$\\
\end{algorithm}

\subsection{The Advanced Algorithm}

For the basic algorithm, although using hyperwedges as samples shows advantages over vertex and hyperedge sampling, it still has some limitations. Firstly, the basic algorithm does not incur any optimization step to estimate specific patterns, therefore, it still needs to find the common vertices among the three hyperedges when classifying hyper-triangles. Furthermore, the basic algorithm is unable to estimate hyper-triangles of specific patterns, and when the sample size is insufficient, hyper-triangles of some patterns are difficult to capture, resulting in an underestimated outcome.


\begin{algorithm}[htb]

\small
\caption{Appro-adv} 
\label{Appro-adv} 
    $G(V,E)\leftarrow$Hypergraph, $\sigma \leftarrow$sampling proportion, $\alpha \leftarrow$number of samples, $\hat{H}\leftarrow$map to store the estimate count of each pattern\\
    $G'(V',E')\leftarrow$sample $\sigma|E|$ hyperedges from $G$ uniformly at random\\
    $\rotatedless^t$, $\rotatedless^c \leftarrow$Preprocess($G'$)\\
    $\alpha_1\leftarrow \frac{|\rotatedless^t|^2\cdot\alpha}{|\rotatedless|^2}$, $\alpha_2\leftarrow \frac{|\rotatedless^c|^2\cdot\alpha}{|\rotatedless|^2}$, $\alpha_3$ and $\alpha_4\leftarrow \frac{|\rotatedless^t||\rotatedless^c|\cdot\alpha}{|\rotatedless|^2}$\\
    
     \For{$n\leftarrow 1 \dots \alpha_1$}{
        $\rotatedless^t_{ij} \leftarrow$sample a hyperwedge from $\rotatedless^t$\\
        $\hat{H}\leftarrow$use \textit{Count-TTT} to count the number of \textit{TTT}-class hyper-triangles that include $\rotatedless^t_{ij}$\\
     }
     \For{$t\leftarrow 9 \dots 20$}{
        $\hat{H}[p]=\hat{H}[p] \cdot \frac{|\rotatedless^t|}{\alpha_1  }\cdot \sigma^{-3}$
     }
    Sample $\alpha_2$, $\alpha_3$, and $\alpha_4$ hyperwedges from $\rotatedless^c$, $\rotatedless^t$, and $\rotatedless^c$\\ 
    $\hat{H}\leftarrow$Estimate the number of hyper-triangles of class \textit{CCC}, \textit{TTC}, and \textit{TCC} following the similar procedure of lines 5-9.\\
     
    {\bf return}\ $\hat{H}$\\
    
\end{algorithm}

To address these issues, we observe that by separately sampling from the lists of intersection-type and inclusion-type hyperwedges, we can independently estimate the number of hyper-triangles for each class. Additionally, during the estimation process, we can incorporate optimization steps specific to each class, thereby enhancing the efficiency of the algorithm. Based on this method, we propose an advanced approximation algorithm that allows for targeted estimation of specific patterns. The detail is presented in Algorithm \ref{Appro-adv}. After extracting a random subgraph from the original graph, we divide the number of samples into four parts according to the method described in line 4. When counting hyper-triangles of the \textit{TTT} class, we randomly sample $\frac{|\rotatedless^t|^2 \cdot \alpha}{|\rotatedless|^2}$ hyperwedges from $\rotatedless^t$, and use the \textit{Count-TTT} algorithm to calculate the number of hyper-triangles that include them (Lines 5-9). For hyper-triangles of the \textit{CCC} class, we sample $\frac{|\rotatedless^c|^2\cdot\alpha}{|\rotatedless|^2}$ hyperwedges from $\rotatedless^c$, and then use \textit{Count-CCC} to count the number of hyper-triangles containing these hyperwedges. Similarly, following the above procedure, we sample $\frac{|\rotatedless^t||\rotatedless^c|\cdot\alpha}{|\rotatedless|^2}$ hyperwedges to estimate the number of the hyper-triangles of the remaining two classes. Note that if our goal is to focus on a specific class, we can sample hyperwedges only for that class, thereby avoiding unnecessary counting.

\begin{lemma}
\label{VarComp}
The time complexity of Appro-adv is: $O(\sum_{\rotatedless_{ij} \in \rotatedless}(|e_i|+|e_j|)+\sum_{i=1}^{\alpha_1}(|\rotatedless_{i\_}^t|\cdot \omega_{ij})+\sum_{i=1}^{\alpha_2}(|\rotatedless_{i\_}^c|)+\sum_{i=1}^{\alpha_3}(|\rotatedless_{i\_}^c|+|\rotatedless_{\_i}^c|)+\sum_{i=1}^{\alpha_4}(|\rotatedless_{i\_}^t|+|\rotatedless_{\_i}^t|))$.
\end{lemma} 

\noindent
{\bf Proof:} In the preprocessing stage, the time complexity of Appro-adv is $O(\sum_{\rotatedless_{ij} \in \rotatedless}(|e_i|+|e_j|))$. After that, it samples $\alpha_1$ hyperwedges and counts the number of hyper-triangles of \textit{TTT} class that contain these hyperwedges through \textit{Count-TTT}. Hence the time complexity is $O(\sum_{i=1}^{\alpha_1}(|\rotatedless_{i\_}^t|\cdot \omega_{ij}))$ (as mentioned in \autoref{Time complexity}). Using the same method, we can obtain the time complexities of the remaining three parts. By combining the preprocessing time with the time of these four parts, we can obtain the total time complexity.

\begin{lemma}
\label{VarComp}
The variances of the results obtained by Appro-adv corresponding to the hyper-triangles of \textit{TTT}, \textit{CCC}, \textit{TCC}, \textit{TTC} class are: $(|\rotatedless^t|-1)\mathcal{T}+\frac{|\rotatedless^t|}{|\rotatedless|}\mathcal{T}'-\mathcal{T}''$, $(|\rotatedless^c|-1)\mathcal{T}+\frac{|\rotatedless^c|}{|\rotatedless|}\mathcal{T}'-\mathcal{T}''$, $\frac{(|\rotatedless^t|-1) |\rotatedless^c|}{|\rotatedless^t|}\mathcal{T}+\frac{|\rotatedless^c|}{|\rotatedless|}\mathcal{T}'-\frac{|\rotatedless^c|}{|\rotatedless^t|}\mathcal{T}''$, $\frac{(|\rotatedless^c|-1) |\rotatedless^t|}{|\rotatedless^c|}\mathcal{T}+\frac{|\rotatedless^t|}{|\rotatedless|}\mathcal{T}'-\frac{|\rotatedless^t|}{|\rotatedless^c|}\mathcal{T}''$ respectively, where $\mathcal{T}=\frac{\alpha H[p]}{|\rotatedless|^2}$, $\mathcal{T}'=\frac{\alpha\gamma_{p}'}{3|\rotatedless|}$ and $\mathcal{T}''=\frac{\alpha\gamma_{p}}{|\rotatedless|^2}$.
\end{lemma} 

\noindent
{\bf Proof:} Based on the proof of Lemma \autoref{variance}, when sampling $\alpha$ hyperwedges, by setting $\delta$ as $|\rotatedless|$, the resulting variance is $\frac{ (|\rotatedless|-1) \alpha H[p]}{|\rotatedless|^2}+\frac{\alpha\gamma_{p}'}{3|\rotatedless|}-\frac{\alpha\gamma_{p}}{|\rotatedless|^2}$. For \textit{Appro-adv}, when estimating the hyper-triangle of the \textit{TTT} class, we sample $\alpha_1$ hyperwedges from $\rotatedless^t$ for estimation where $\alpha_1 = \frac{|\rotatedless^t|^2\cdot\alpha}{|\rotatedless|^2}$. Hence, we replace $\rotatedless$ and $\alpha$ with $\rotatedless^t$ and $\alpha_1$, respectively. After simplification, we obtain the variance of the \textit{TTT} class: $\frac{ (|\rotatedless^t|-1) \alpha H[p]}{|\rotatedless|^2} + \frac{|\rotatedless^t|}{|\rotatedless|}\frac{\alpha \gamma_{p}'}{3|\rotatedless|} - \frac{\alpha \gamma_{p}}{|\rotatedless|^2}$. Using the same method, we can obtain the variance of the remaining three classes.

\vspace{1.5mm}\noindent
{\bf Comparing the two approximate algorithms.} As mentioned earlier, the second term in the variance is the largest term. The variance of \textit{Appro-bs} is given by $\frac{ (|\rotatedless|-1) \alpha H[p]}{|\rotatedless|^2}+\frac{\alpha\gamma_{p}'}{3|\rotatedless|}-\frac{\alpha\gamma_{p}}{|\rotatedless|^2}$. Obviously, its second term is greater than the second term in all the variances from Lemma \autoref{VarComp}, leading to a larger error in the results obtained through \textit{Appro-bs}. Additionally, apart from accuracy, since \textit{Appro-adv} uses methods from exact algorithms to compute the number of hyper-triangles, it requires less time when the sample sizes are the same.

\section{The Fine-grained Hypergraph Clustering Coefficient}
\label{Clustering Coefficient}

\begin{figure*}[t]
	\begin{center}
            \subfigure[Datasets from different domains]{
			\includegraphics[scale=0.15]{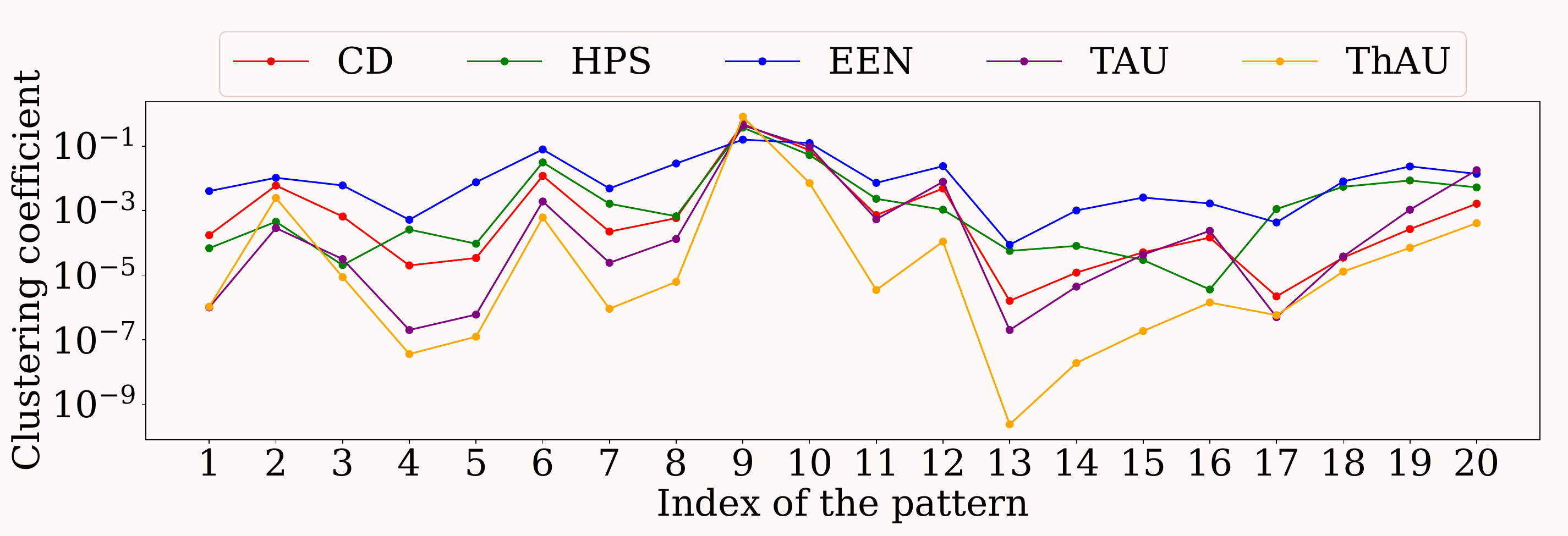}     
		}\hspace{5mm}
		\subfigure[Datasets from same domain]{ 
			\includegraphics[scale=0.15]{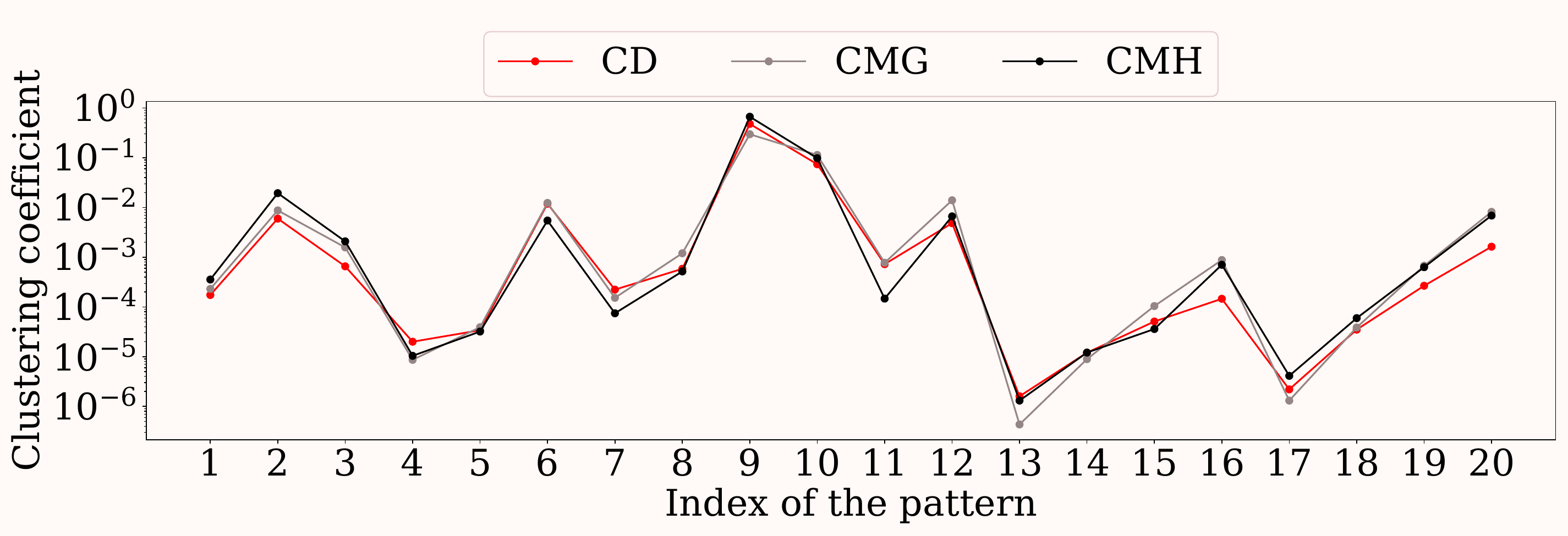}   
		}\vspace{-1mm}
	\end{center}
        \vspace{-2mm}
	\caption{Clustering coefficient comparison}
    
    \label{Clustering coefficient Comparison}
    \vspace{-2mm}
\end{figure*}

In graph theory, the clustering coefficient is a metric used to measure the tendency of nodes within a graph to form clusters or closely-knit groups. Based on the definition of clustering coefficient in general graphs \cite{seshadhri2014wedge, green2013faster}, existing studies \cite{estrada2006subgraph, aksoy2020hypernetwork} propose the definition of clustering coefficient for hypergraphs, which is $3 \times |\customtriangle| / |\opentopsquare|$, where $\customtriangle$ is the set of hyper-triangles and $\opentopsquare$ is the set of open hyper-triangles (i.e., a hyper-triangle in which two hyperedges are disconnected). 

While the current definition can be used to measure the localized closeness and redundancy of the hypergraphs, it still has some limitations. Since the semantic meanings vary across different hyper-triangle patterns, there are cases where we prioritize the information conveyed by specific patterns. Therefore, it is crucial to distinguish between different patterns. Motivated by this, we propose a fine-grained hypergraph clustering coefficient, which enables us to adjust the proportion of each pattern according to the domains of the datasets. Note that by setting $\epsilon_p=1$ for each pattern $p$, our model can be transformed into the existing model.

\begin{definition}
\label{cc}
[{\bf Fine-grained hypergraph clustering coefficient}] Given a hypergraph \( G = (V, E) \), the fine-grained clustering coefficient of $G$ is given by $\frac{3\times\sum_{p=1}^{20}\epsilon_p |\customtriangle_p|}{|\opentopsquare|}$, where $\epsilon_p \in [0,1]$ is a parameter for the hyper-triangle pattern $p$, $|\customtriangle_p|$ is the number of hyper-triangles of pattern $p$ in $G$ and $|\opentopsquare|$ is the number of open hyper-triangles (i.e., a hyper-triangle in which two hyperedges are disconnected) in $G$.

\end{definition} 

To clearly compare the distribution of different hyper-triangle patterns across datasets from various domains, we calculate the hypergraph clustering coefficient for each pattern in real-world datasets by setting the parameter of the current pattern to 1 and all others to 0. The results are displayed in \autoref{Clustering coefficient Comparison}. From these figures, we can observe that the distribution frequency of patterns varies when the data comes from different domains. However, when the data originates from the same domain, the distribution frequency of patterns remains consistent. This indicates that data from specific domains, due to its inherent properties, tends to form certain specific hyper-triangle patterns. In some cases, we may prefer to use specific patterns to reflect the inherent properties of datasets. Hence, we develop the fine-grained clustering coefficient for hypergraphs, which can be flexibly adjusted to more intuitively display the information contained in different patterns.


\section{Experiments}
\label{Experiments}

In this section, we evaluate the effectiveness and efficiency of our proposed algorithms. 

\subsection{Experimental Settings}

\begin{table}[htb]
    
    \centering
    \resizebox{0.43\textwidth}{!}{%
    \begin{tabular}{|l|l|l|l|l|l|}
        \hline
        \multicolumn{1}{|c|}{\cellcolor{gray!25}\textbf{Dataset}} &
        \multicolumn{1}{c|}{\cellcolor{gray!25}\textbf{|V|}} & 
        \multicolumn{1}{c|}{\cellcolor{gray!25}\textbf{|E|}} & 
        \multicolumn{1}{c|}{\cellcolor{gray!25}\textbf{$|e|_{max}$}}  & 
        \multicolumn{1}{c|}{\cellcolor{gray!25}\textbf{$|e|_{avg}$}} & 
        \multicolumn{1}{c|}{\cellcolor{gray!25}\textbf{|$\rotatedless$|}} \\
        \hline
         HPS & 327 & 7,818 & 5 & 2.3 & 593K \\
        \hline 
         CMH & 1,014,734 & 895,439 & 25 & 1.5 & 1.7M \\
        \hline 
         CPS & 242 & 12704 & 5 & 2.4 & 2.2M \\
        \hline
         EEU & 998 & 25,027 & 25 & 3.4 & 8.3M \\
        \hline
         ThAU & 125,602 & 166,999 & 14 & 1.9 & 21.6M \\
        \hline
         CMG & 1,256,385 & 1,203,895  & 25 & 3.1 & 37.6M \\
        \hline
         EEN & 143 & 1512 & 18 & 3.0 & 87.8M \\
        \hline
         CD &1,924,991 & 2,466,792 & 25 & 2.9 & 125M \\
        \hline
         TAU & 3029 & 147,222  & 5 & 3.4 & 564M \\
        \hline
         ThM& 176,445 & 595,749  & 21 & 2.4 & 647M \\
        \hline
         TMS & 1629 & 170,476 & 5 & 3.5 & 913M \\
        \hline
    \end{tabular}%
    }
    \vspace{2mm}
    \caption{Summary of Datasets}
    \label{Datasetl} 
    \vspace{-9mm}
\end{table}

\noindent
{\bf Datasets.} In the experiments, we use the following 11 datasets. The co-authorship \cite{sinha2015overview} domain includes four datasets (coauthor-DBLP (CD), coauthor-geology (CMG), and coauthor-history (CMH)). In these datasets, vertices represent authors, and hyperedges represent publications, with the authors being the vertices included in the hyperedges. The contact domain comprises two datasets (contact-primary (CPS) \cite{stehle2011high} and contact-high (HPS) \cite{mastrandrea2015contact}). In these datasets, vertices represent individuals, while hyperedges represent relationships between individuals. Email-EU (EEU) \cite{leskovec2005graphs, yin2017local} and email-Enron (EEN) \cite{klimt2004enron} are two datasets belong to the email domain, where vertices represent email entities, and hyperedges represent either senders or receivers of the emails. The tags domain and threads domain each consist of two datasets. They are tags-ubuntu (TAU), tags-math (TMS), threads-ubuntu (ThAU), and threads-math (ThM). In the datasets from the tags domain, vertices represent tags, and hyperedges represent posts containing these tags. In the threads domain, vertices represent users, and hyperedges represent threads in which users are involved. These datasets are sourced from \cite{benson2018simplicial}. Due to the property of hypergraphs, where any two hyperedges cannot contain the same set of vertices, we performed deduplication on the original data. \autoref{Datasetl} provides details about the datasets.

\noindent
{\bf Algorithms.} In the experiments, we use \textit{Exact-bs} as the baseline for the exact algorithm and compare its performance with the following five algorithms: \textit{Count-CCC} (for pattern 1), \textit{Count-TCC} (for patterns 2-5), \textit{Count-TTC} (for patterns 6-8), \textit{Count-TTT} (for patterns 9-20), and \textit{Exact-adv}. Here \textit{Exact-adv} is to compute hyper-triangles of all the patterns by combining the search processes of \textit{Count-TTT} and \textit{Count-TTC}, as well as \textit{Count-TCC} and \textit{Count-CCC}, respectively. Subsequently, we conduct a comparison between \textit{Count-TTT}, \textit{Count-DenseTTT} (for patterns 9-16) and \textit{Count-SparseTTT} (for patterns 17-20) in terms of the time required to find \textit{TTT}-class hyper-triangles. For the approximation algorithms, we examine the performance of \textit{Appro-bs} and \textit{Appro-adv}. Similarly, in the parallel algorithms section, we compare the performance between \textit{Par-bs} and \textit{Par-adv}. All the algorithms are implemented in C++, and all experiments are conducted on a Linux machine with an Intel(R) Xeon(R) Platinum 8260L CPU at 2.30 GHz and 256GB of memory.

\subsection{Case Study}

\begin{figure}[htb]
\begin{minipage}{0.26\textwidth}
\centering
\resizebox{\textwidth}{!}{%
    \small
    \begin{tabular}{|c c c|}
        \hline
        \multicolumn{3}{|c|}{Publications (Authors)}                   \\
        \hline
        $e_1$: (1, 2, 3) &  $e_2$: (3, 4, 5) & $e_3$: (2, 3, 4, 6) \\
        $e_4$: (2, 3, 4, 7) & $e_5$: (4, 8, 9, 10) & \\
        \hline
    \end{tabular}
}\vspace{1mm}
\resizebox{0.99\textwidth}{!}{%
    \small
    \begin{tabular}{|c c c|}
        \hline
        \multicolumn{3}{|c|}{Hyper-triangles (Pattern)}                    \\
        \hline
        $\{e_1, e_2, e_3\}$: 12 & $\{e_2, e_3, e_4\}$: 10 & $\{e_3, e_4, e_5\}$: 10\\
        \hline
    \end{tabular}
}
\end{minipage}\hspace{1mm}
\begin{minipage}{0.20\textwidth}
\centering
    \includegraphics[scale=0.35]{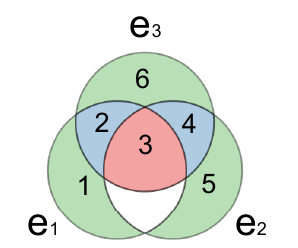}
    \hspace{-3mm}
    \includegraphics[scale=0.35]{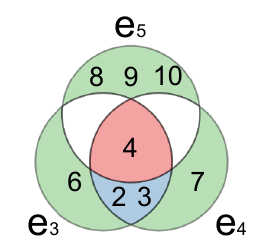}
\end{minipage}
\vspace{0mm}
\caption{Real data for co-authorship relations}

\label{casestudyprocess}

\end{figure}
\vspace{-2mm}

\begin{figure}[htb]
\noindent
\begin{minipage}{0.23\textwidth}
\centering
\resizebox{\textwidth}{!}{%
    \small
    \begin{tabular}{|c c c|}
        \hline
        \multicolumn{3}{|c|}{Email account (Senders or Receivers)} \\
        \hline
        \multicolumn{3}{|c|}{$e_1$: (1, 29, 41, 51, 62, 65, 97, 107, 133)} \\
        $e_2$: (1, 51) & $e_3$: (1, 133) & $e_4$: (29, 97) \\
        $e_5$: (41, 97) & $e_6$: (29, 65) & \\
        \hline
    \end{tabular}
}\vspace{1mm}
\resizebox{0.99\textwidth}{!}{%
    \small
    \begin{tabular}{|c c c|}
        \hline
        \multicolumn{3}{|c|}{Hyper-triangles (Pattern)}                    \\
        \hline
        $\{e_1, e_2, e_3\}$: 5 & $\{e_1, e_4, e_5\}$: 5 & $\{e_1, e_4, e_6\}$: 5\\
        \hline
    \end{tabular}
}
\end{minipage}\hspace{-1mm}
\begin{minipage}{0.16\textwidth}
    \includegraphics[scale=0.62]{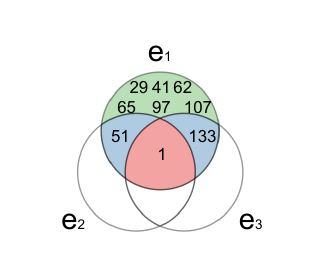}
\end{minipage}

\vspace{-3mm}
\caption{Real data for email correspondence relations}

\label{casestudyprocess2}
\vspace{0.5mm}
\end{figure}

\begin{figure*}[htb]
	\begin{center}
            \subfigure[]{
			\label{total runtime}
			
			\includegraphics[scale=0.178]{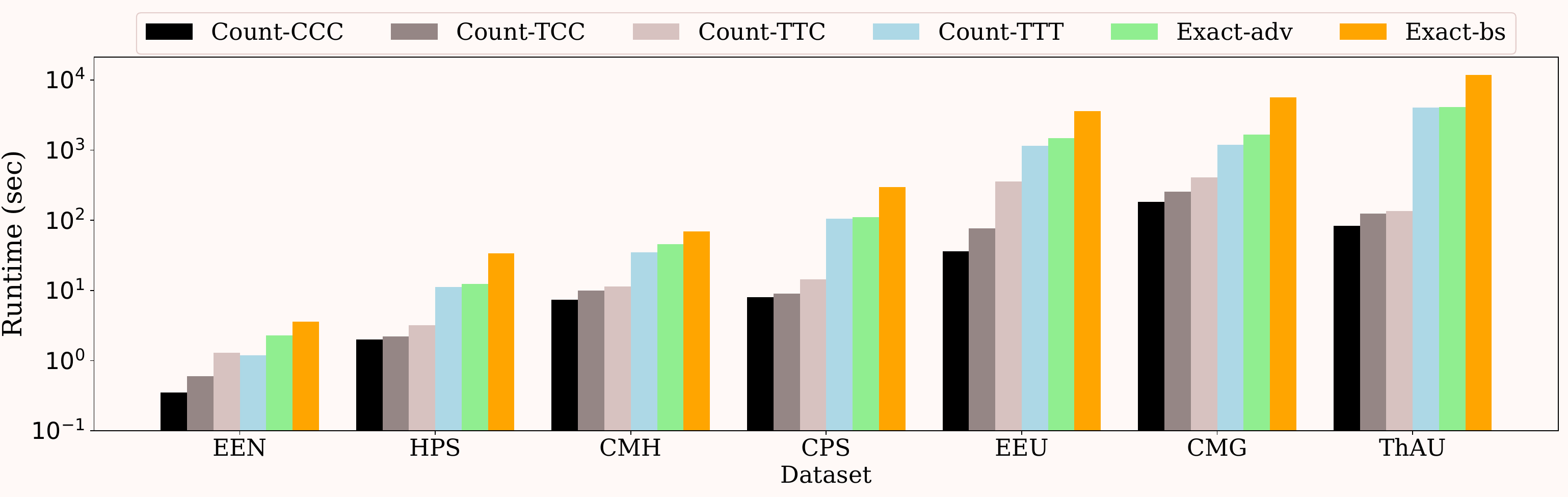}      
		}\hspace{4mm}
		\subfigure[]{
			\label{TTT runtime}
                
			     \includegraphics[scale=0.178]{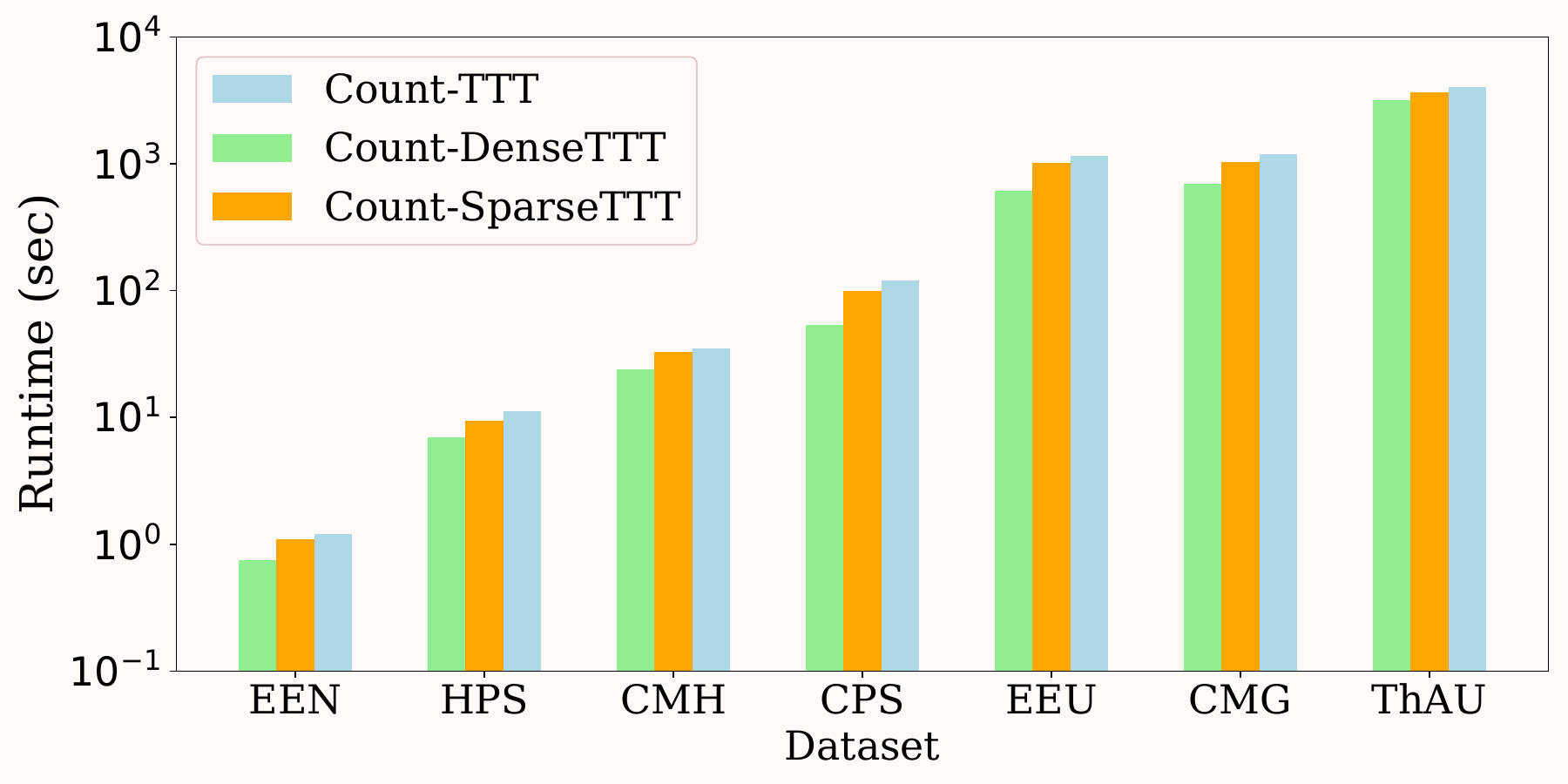} 
		}
    
	\end{center}
        \vspace{-4mm}
	\caption{Runtime for exact algorithms.}
    
    \label{exact data}
    \vspace{-3mm}
\end{figure*}

Different hyper-triangle patterns encapsulate distinct meanings, leading to variations in their frequency distribution across various datasets. These variations reveal underlying design principles specific to each dataset. As analyzed in \cite{lee2020hypergraph}, pattern 12 is more prevalent in the co-authorship domain, whereas pattern 10 frequently appears in email correspondence domain. To understand the reasons behind these distributions, we conduct case studies that identify and analyze typical hyper-triangles in these two domains. 

In the co-authorship domain, we use the CMH dataset to analyze the hyper-triangle patterns. \autoref{casestudyprocess} displays ten authors involved in three hyper-triangles, along with the five publications. The dataset shows that Authors 1-7 have collaborated over twenty times, while Authors 8-10 have more than ten collaborations, indicating they are likely from two different groups. Through hyper-triangle $\{e_1, e_2, e_3\}$, we can see that all three publications are centered around Author 3, and Author 3 maintains a stable co-authorship with Authors 2 and 4. Authors 1, 5, and 6, as the first authors of the three publications, do not have direct connections to each other. In fact, most hyper-triangles of pattern 12 display a similar structure to the one described above, as completing a publication usually involves multiple contributors arranged in a hierarchical structure. A senior researcher often initiates the project (Author 3), followed by contributions from young scholars (Authors 2 and 4). Finally, several student participants also contribute to the publication (Authors 1, 5 and 6). For the collaborative relationships between different groups (e.g., the hyper-triangle $\{e_3, e_4, e_5\}$), we can observe that $e_3$ and $e_4$, which are publications from the same group, share three coauthors. In contrast, the publication $e_5$, originating from another group, has only one coauthor in common with $e_3$ and $e_4$. This suggests that inter-group collaborations often depend on specific authors (in this case, Author 4). As a result, the hyper-triangles formed are more likely to align with pattern 10. This reasoning indicates that co-authorship datasets are naturally predisposed to contain a higher number of hyper-triangles of pattern 12.

In the email correspondence domain, \autoref{casestudyprocess2} show three hyper-triangles from EEN, where hyperedges represent email accounts and the vertices correspond to accounts that have sent or received emails from those accounts. It can clearly be seen that these hyper-triangles contain hyperedge $e_1$, and $e_1$ encompasses the remaining five hyperedges. This is because, in reality, some organizations have "organizer" or admin account that frequently sends and receives emails to and from its member accounts. Consequently, the member accounts tend to interact predominantly with the 'organizer' or admin accounts. Thus, compared to other domains, datasets in the email correspondence domain are more prone to forming hyper-triangles of pattern 5.


\subsection{Performance Evaluations}

{\bf Evaluating exact algorithms.} In \autoref{exact data}(a), we present the comparison of exact algorithms in 7 datasets. Through \autoref{exact data}(a), it can be observed that \textit{Count-CCC}, \textit{Count-TCC}, and \textit{Count-TTC} significantly outperforms \textit{Exact-bs} by up to 140$\times$, 90$\times$ and 80$\times$, respectively. This is because these three algorithms can avoid enumerating hyper-triangles of the \textit{TTT} class and, compared to \textit{Exact-bs}, do not require additional time to determine the pattern of the hyper-triangle. While \textit{Count-TTT} requires more time than the other algorithms, it is still at least 3$\times$ faster than \textit{Exact-bs}. In addition, \textit{Exact-adv} is capable of computing all hyper-triangles in a shorter time compared to \textit{Exact-bs}, which achieves at least 3$\times$ faster than \textit{Exact-bs} in the large datasets. These findings affirm the efficacy of our proposed techniques in reducing the redundant enumeration of vertices within hyperedges. 

\vspace{1.5mm}\noindent{\bf Evaluating exact algorithms for \textit{TTT} class.} For the three algorithms targeting the \textit{TTT} class: \textit{Count-DenseTTT}, \textit{Count-SparseTTT} and \textit{Count-TTT}, we also use the aforementioned 7 datasets to compare their differences. The results are displayed in \autoref{exact data}(b). It can be observed that \textit{Count-DenseTTT} has the least runtime. For \textit{Count-DenseTTT}, it avoids traversing open hyper-triangles, thereby achieving about 2$\times$ faster than \textit{Count-TTT}. For \textit{Count-SparseTTT}, since finding hyper-triangles of the \textit{SparseTTT} class still requires time to filter open hyper-triangles, the runtime is about $1.5\times$ faster than \textit{Count-TTT}.


\begin{figure}[htb]
    \begin{center}
        \centering
        \includegraphics[scale=0.18]{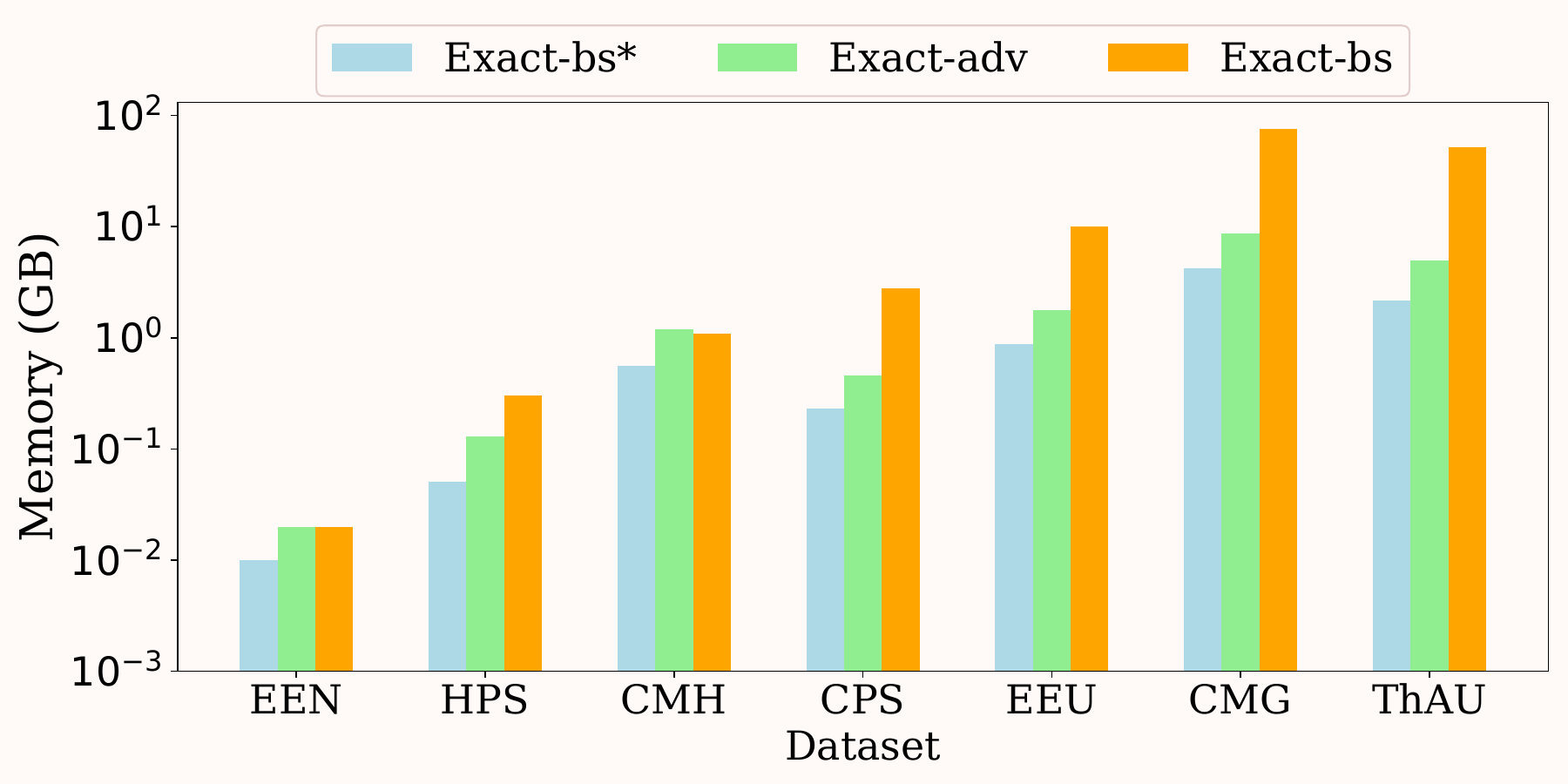} 
	\end{center}
	\vspace{-3mm}
	\caption{The memory required for exact algorithms.} 
	\label{Space}
	\vspace{-2mm}
\end{figure}

\vspace{1.5mm}\noindent{\bf Evaluating the space for the exact algorithms.} 
Since the source code of \textit{Exact-bs} \cite{lee2020hypergraph} additionally stores the 2-hop neighbors of each hyperedge, leading to extra space consumption, we improve its code to store only the 1-hop neighbors of each hyperedge without affect the performance (denoted as \textit{Exact-bs*}). \autoref{Space} examines the memory usage of \textit{Exact-bs}, \textit{Exact-bs*} and \textit{Exact-adv}. From the results, we observe that \textit{Exact-adv} consistently consumes about 2$\times$ the space of \textit{Exact-bs*}. This means that compared to the baseline algorithm, the proposed algorithms can find all hyper-triangles with at least 3$\times$ the efficiency while only using 2$\times$ the space. Besides, if the goal of users is to find hyper-triangles of a specific pattern, the proposed algorithms can even achieve up to 140$\times$ faster than the baseline algorithm without increasing space consumption.

\vspace{1.5mm}\noindent{\bf Evaluating approximate algorithms.} Here we compare the performance of \textit{Appro-bs} and \textit{Appro-adv}. We use $\sum_{p=1}^{20}\frac{|H[p]-\hat{H}[p]| }{20\times H[p]}$ as the metric to measure the accuracy of the algorithms where $H[p]$ and $\hat{H}[p]$ are the exact and estimated results of the hyper-triangles of pattern $p$, respectively. We first randomly extract 20\% of the hyperedges from the original graph and then perform a comparison on this induced subgraph. By comparing the accuracy of the results obtained by the two algorithms within the same time period, we can determine the differences between the two algorithms. 

\begin{figure}[t]
    \begin{center}
            \subfigure[TAU]{
			\label{Appro-TAU}
			\centering
			\includegraphics[scale=0.073]{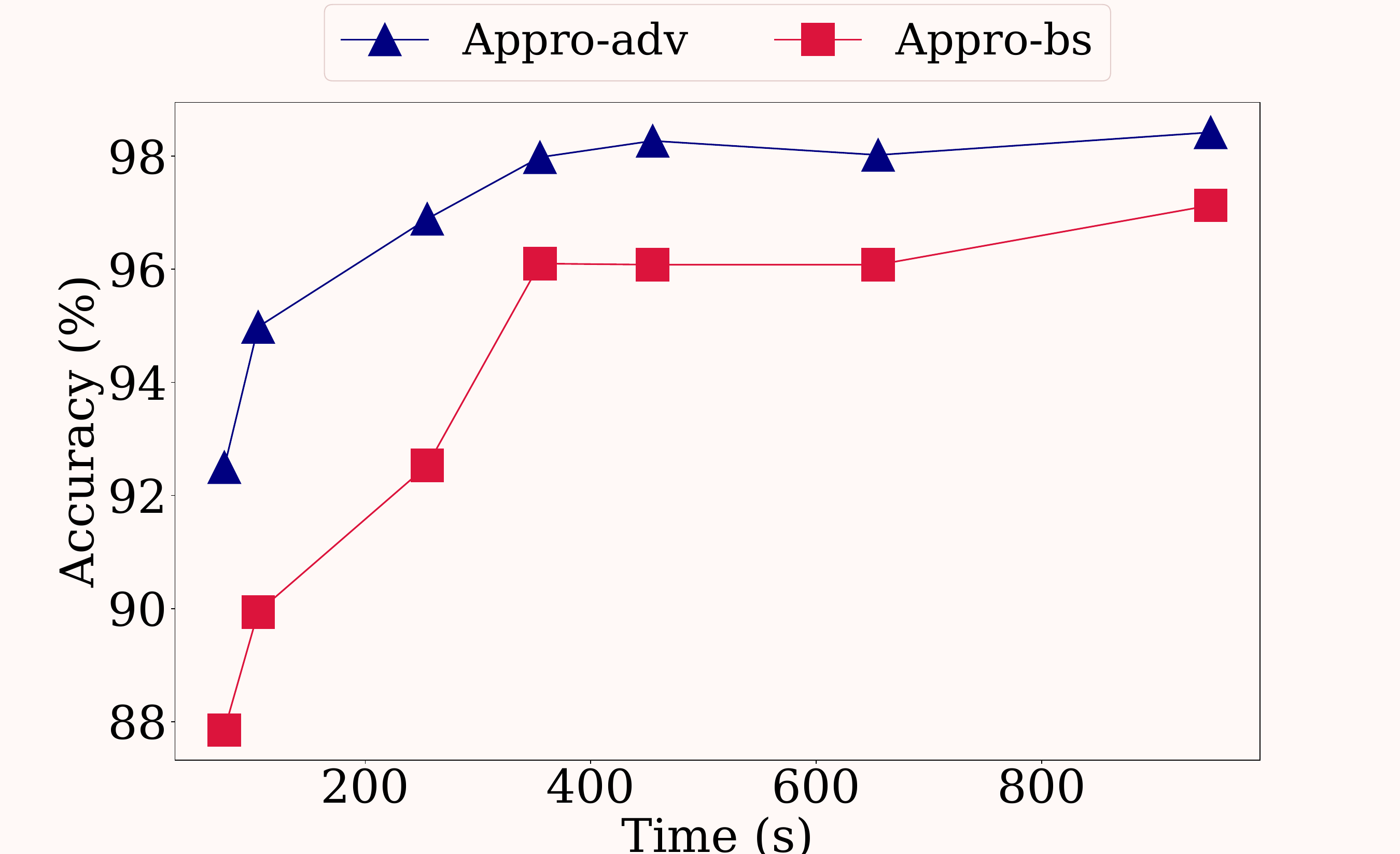} 
		}\hspace{-2mm}
		\subfigure[ThM]{
			\label{Appro-ThM}
			\centering
			\includegraphics[scale=0.073]{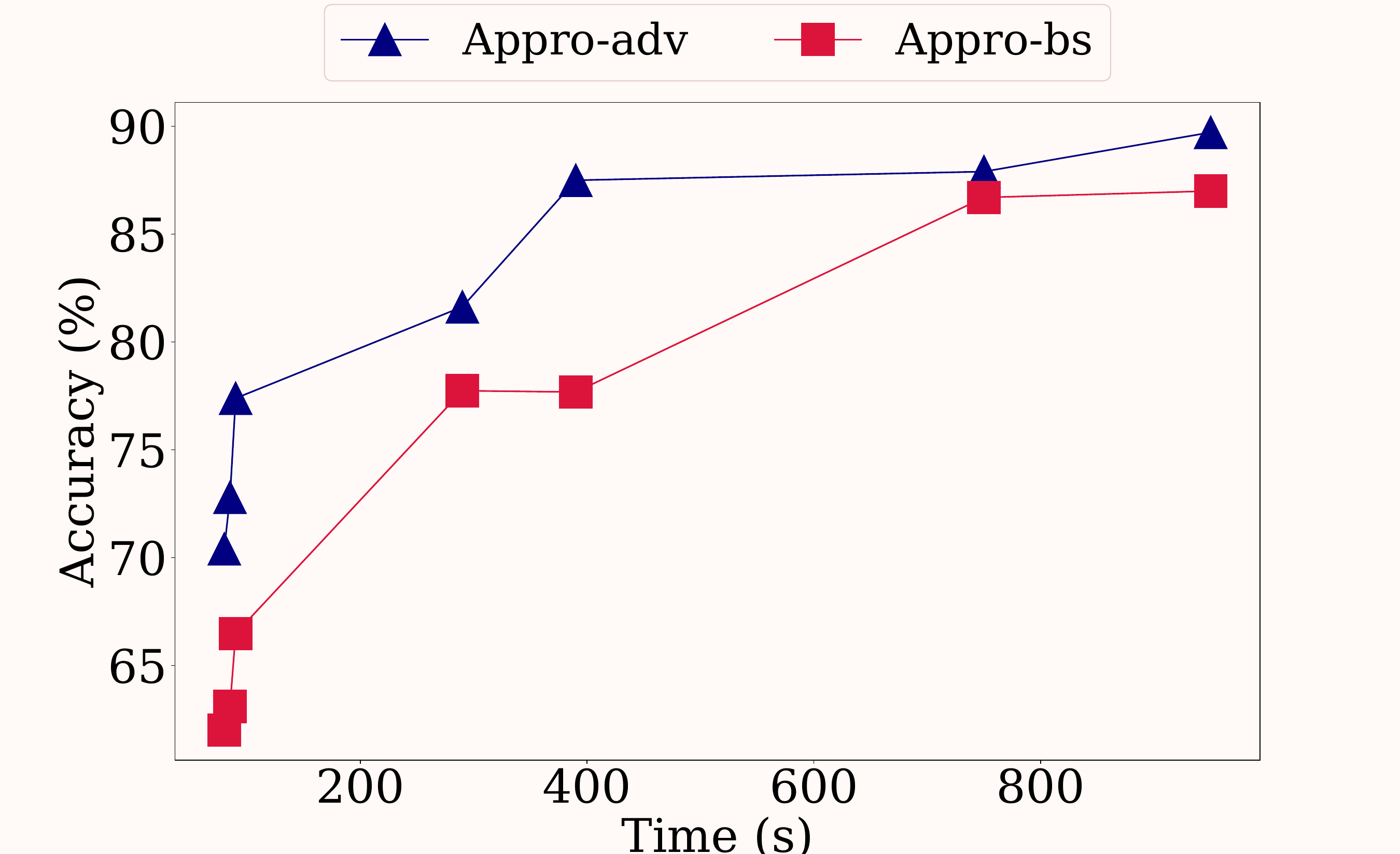}     
		}\vspace{-2.5mm}
            \subfigure[TMS]{
			\label{Appro-TMS}
			\centering
			\includegraphics[scale=0.073]{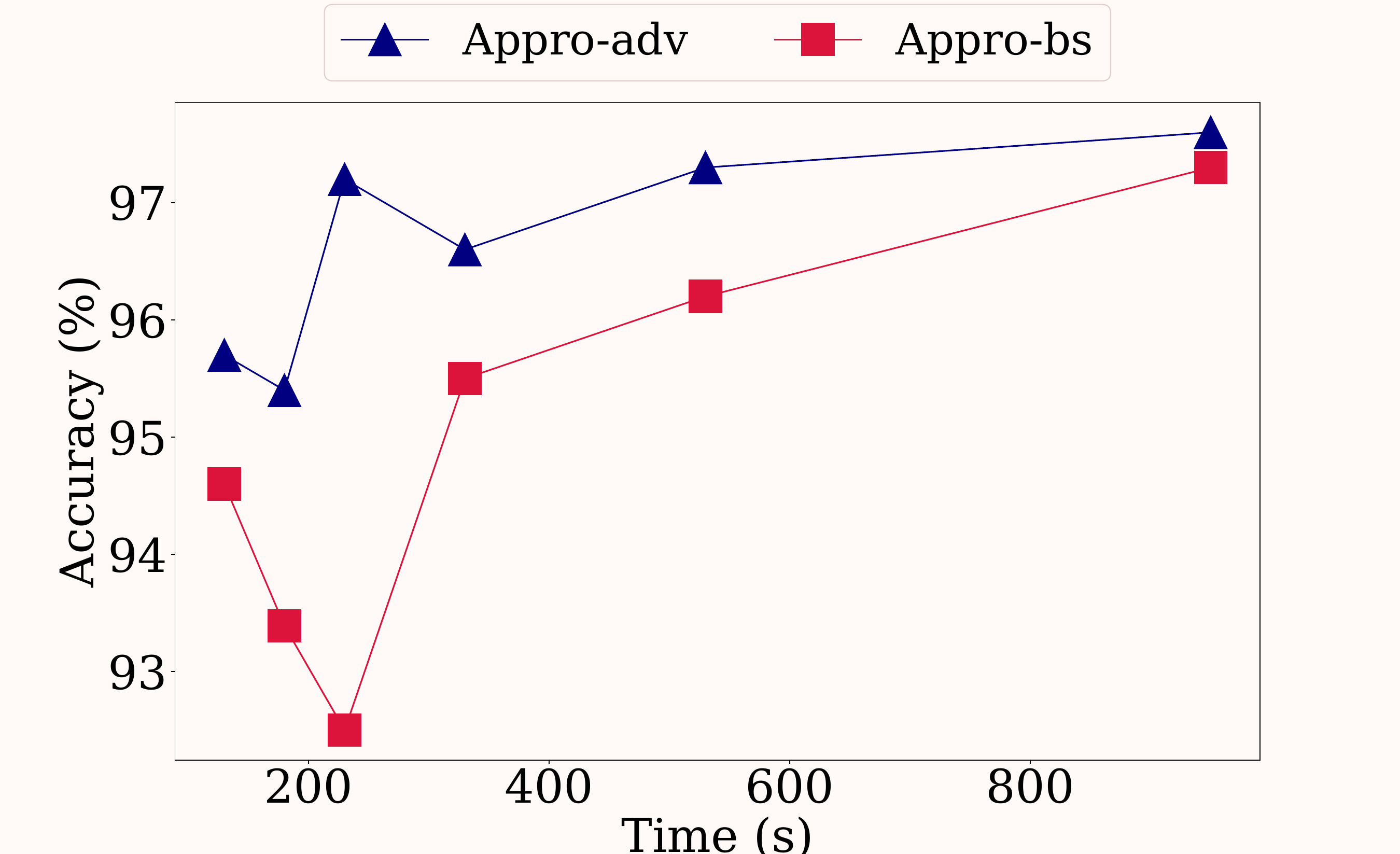}     
		}\hspace{-2mm}
            \subfigure[CD]{
			\label{Appro-CD}
			\centering
			\includegraphics[scale=0.073]{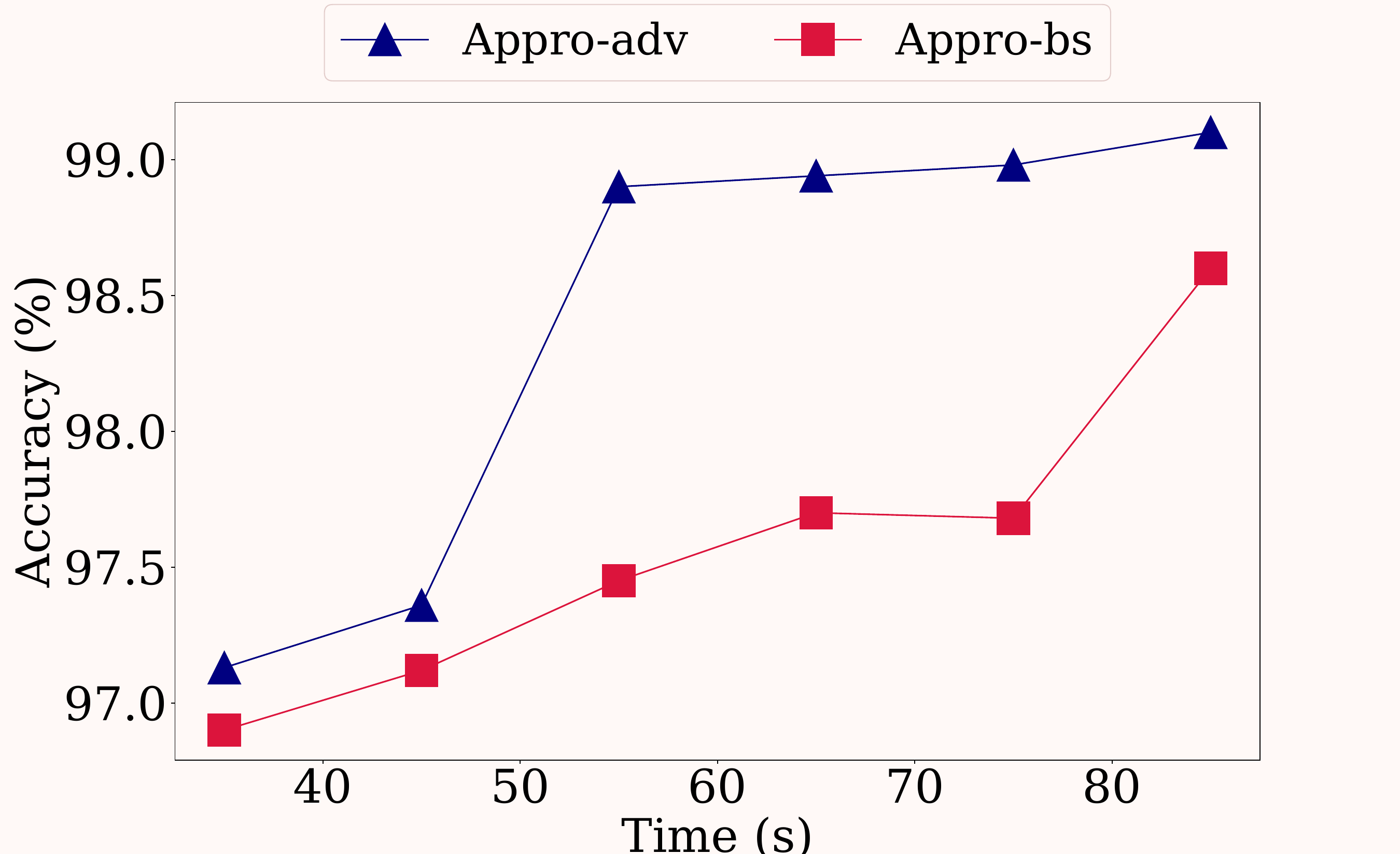}     
		}
	\end{center}
	\vspace{-3mm}
	\caption{Approximate algorithms' result, $\sigma=0.2$.} 
	\label{Fully approximation result}
	\vspace{-4mm}
\end{figure}

\autoref{Fully approximation result} shows the performance of the two approximation algorithms across four datasets: TAU, ThM, TMS, and CD. Due to their large size, we are unable to get the exact number of hyper-triangles in the original graph. Therefore, we only compute the exact number of hyper-triangles in the subgraph, and based on this, we compare the relative accuracy of the two approximation algorithms. It can be observed that for the first three datasets, when the estimation time is less than 400 seconds, the accuracy of \textit{Appro-adv} is at least 5\% higher than \textit{Appro-bs}. Only when the estimation time exceeds 800 seconds does the gap between them narrow. For the CD dataset, the gap is also very significant when the estimation time is less than 80 seconds. The experimental results show that the accuracy of \textit{Appro-adv} is at most 3\% higher than \textit{Appro-bs}. Furthermore, the superiority of \textit{Appro-adv} is especially pronounced when the estimation time is brief. These findings clearly confirm the superiority of \textit{Appro-adv} over \textit{Appro-bs}. Additionally, we can observe that for the TAU, ThM, TMS, and CD datasets, \textit{Appro-adv} can estimate the number of hyper-triangles within 400 seconds with an accuracy close to 90\%. In contrast, the exact algorithm requires more than 10,000 seconds to obtain accurate results. This indicates that when users prioritize efficiency over accuracy, the approximation algorithm, compared to the exact algorithm, has a clear advantage.

\begin{figure}[t]
 
    \begin{center}
            
		\subfigure[EEU]{
			\label{par-TAU}
			\centering
			\includegraphics[scale=0.073]{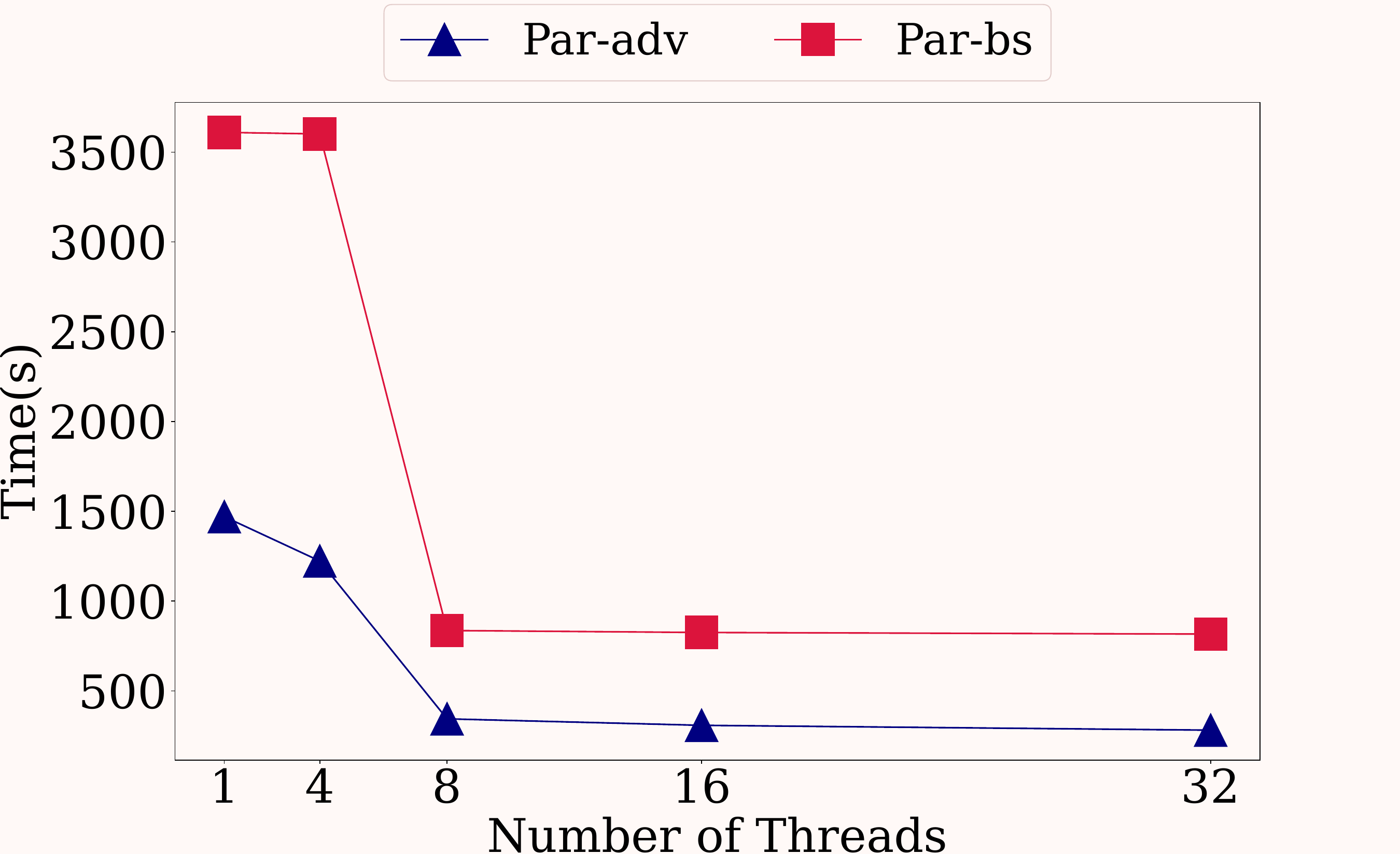} 
		}\hspace{-2mm}
		\subfigure[CPS]{
			\label{par-CPS}
			\centering
			\includegraphics[scale=0.073]{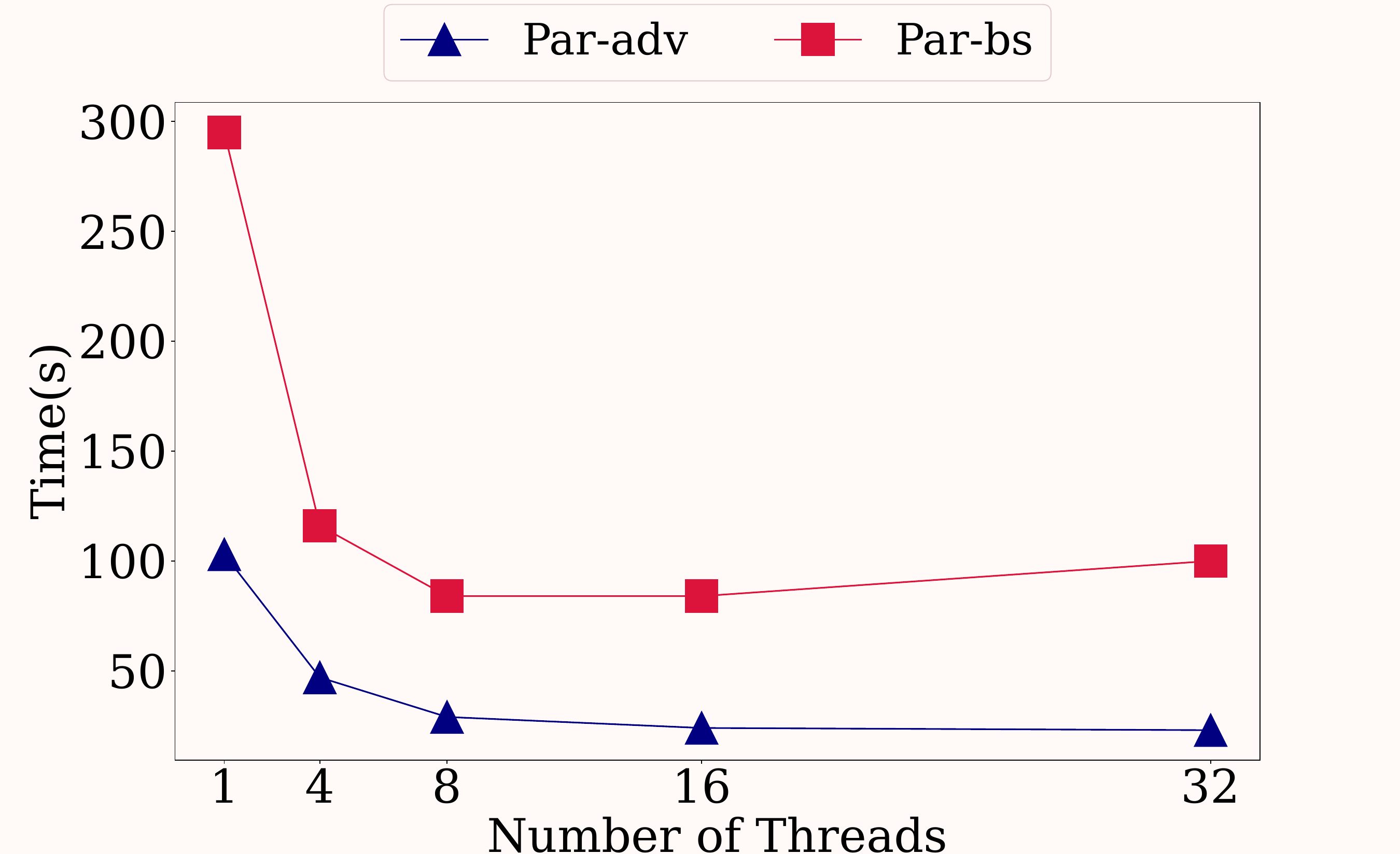}     
		}\vspace{-2.5mm}
            \subfigure[ThAU]{
			\label{par-ThAU}
			\centering
			\includegraphics[scale=0.073]{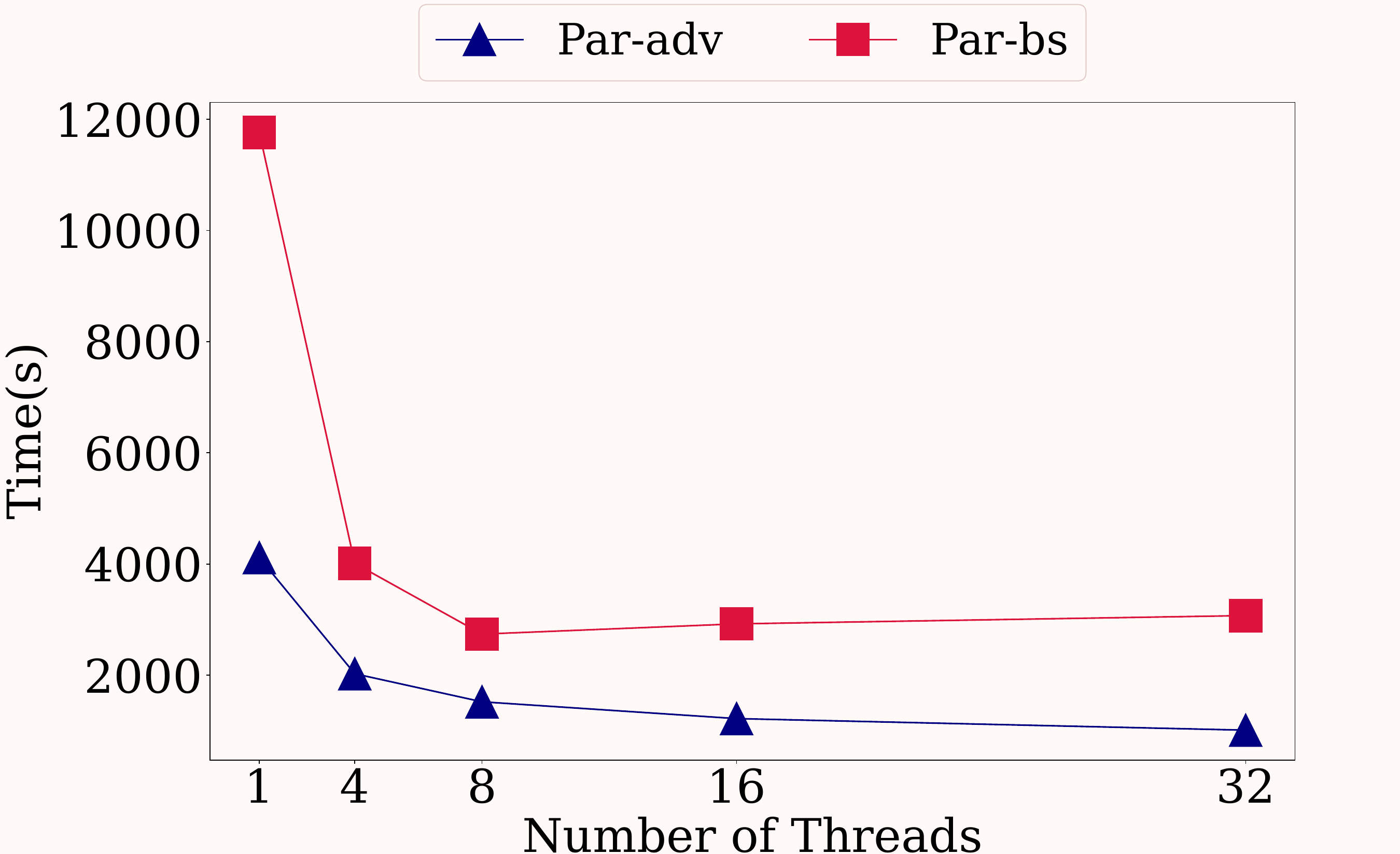}     
		}\hspace{-2mm}
            \subfigure[CMH]{
			\label{par-CMH}
			\centering
			\includegraphics[scale=0.073]{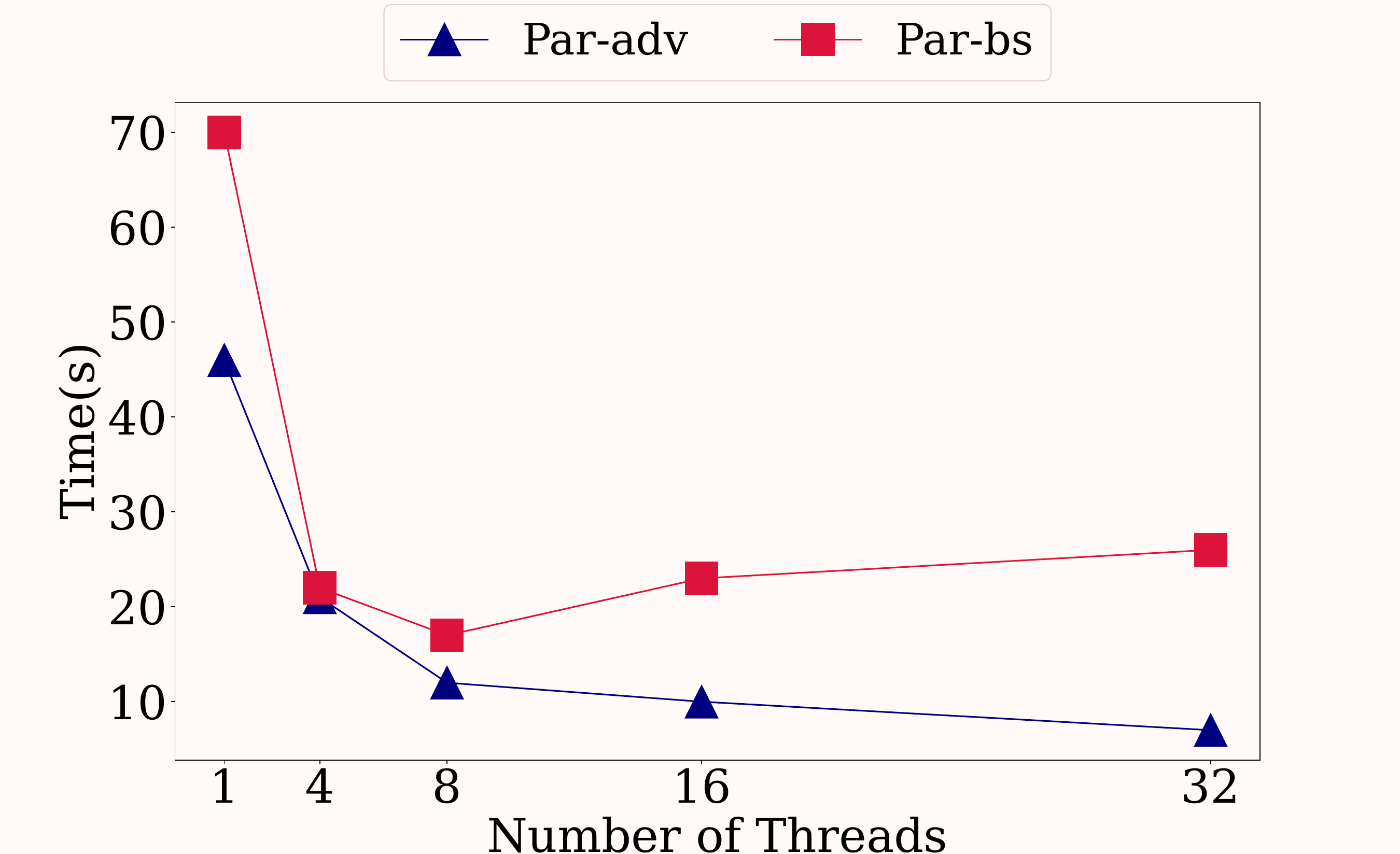}     
		}
	\end{center}
	\vspace{-4mm}
	\caption{Parallel algorithms' result.} 
	\label{parallel result}
	\vspace{-4.5mm}
\end{figure}

\vspace{1.5mm}\noindent{\bf Evaluating parallel algorithms.} In this section, we compare the runtime of \textit{Par-bs}. and \textit{Par-adv} under different numbers of threads in 6 datasets. The experimental results are presented in \autoref{parallel result}. 
For \textit{Par-bs}, across all datasets, the runtime decreases gradually with an increasing number of threads. However, when the number of threads exceeds 16, the runtime begins to increase. In contrast, for \textit{Par-adv}, as the number of threads increases, the runtime of the algorithm first decreases rapidly and then stabilizes. When we use 32 threads, \textit{Par-adv} achieves at least 3$\times$ faster than \textit{Par-bs}.


\section{Related work}
\label{Related work}

\noindent
\textit{Triangle counting.} In general graphs, triangle represents the smallest non-trivial cohesive structure, and there are extensive studies on counting triangles in the literature \cite{berry2011listing, chiba1985arboricity, chu2011triangle, latapy2008main, schank2005finding, yu2020aot, ravichandran2023fast, hu2021accelerating, ahmed2017sampling, ahmed2014graph, pavan2013counting, lim2015mascot}. In \cite{chiba1985arboricity}, Chiba and Nishizeki propose an enumeration-based algorithm to find triangles that orders the vertices by degree and processes each edge only once, using its lower-degree vertex. Some existing works \cite{ahmed2017sampling, ahmed2014graph} focus on edge sampling strategies for estimating triangle counts in a graph stream. Building on this, \cite{pavan2013counting, lim2015mascot} propose a wedge-based sampling method to estimate the number of triangles. In \cite{yu2020aot}, an orientation-based algorithm is proposed to improve the efficiency of finding triangles. \cite{hu2021accelerating} proposes a novel lightweight graph preprocessing method for triangle counting using GPU. 
However, existing triangle counting techniques cannot be directly used for hyper-triangle counting since there are significant structural differences between triangle and hyper-triangle.


\vspace{1mm}\noindent\textit{Hypergraph analysis.} Hypergraphs naturally represent group interactions and are widely used in various fields such as social networks \cite{li2013link, yang2019revisiting}, bioinformatics \cite{hwang2008learning}, recommendation systems \cite{li2013news, bu2010music}, and LLM \cite{feng2024beyond, mao2025survey}. Existing research extensively explores motifs in hypergraphs \cite{zhou2006learning, bretto2013hypergraph}. Hypergraph clustering coefficients are studied in \cite{amburg2020clustering, yoon2020much} to express the overall connectivity among hyperedges and can be used to measure the localized closeness and redundancy of hypergraphs. \cite{zhang2023efficiently} proposes sampling-and-estimating frameworks for counting three types of triangles over hypergraph streams. In \cite{lee2020hypergraph, kook2020evolution}, researchers focus on finding the connectivity patterns among three hyperedges and reveal that the distribution frequency of these patterns varies across different data domains. Although the algorithms in \cite{lee2020hypergraph, kook2020evolution} can be adopted to solve the hyper-triangle computation problem, our proposed techniques outperform existing solutions as validated in our experiments.

\section{Conclusion}
\label{Conclusion}

In this paper, we investigate the hyper-triangle computing problem. We present a two-step framework that efficiently searches for hyper-triangles of specific patterns, supplemented by a parallel version to handle large datasets. Additionally, we propose approximation algorithms for counting the number of hyper-triangles in hypergraphs. Furthermore, we introduce a fine-grained model of the hypergraph clustering coefficient, offering greater flexibility for application across diverse datasets. Through extensive experiments on real datasets, we demonstrate the significant superiority of our algorithms over state-of-the-art approaches. 

\noindent
In the future, we will explore how to extend our approach to search for larger hypergraph patterns. Firstly, since hyper-triangles are a specific case of hyper-cliques, the proposed algorithms can be adapted to identify hyper-cliques by counting the number of hyper-triangles associated with each hyperedge, enabling us to filter hyperedges and narrow the search range. Additionally, in general graphs, triangles serve as fundamental building blocks for the truss model. By counting triangles, we can derive truss structures that facilitate community detection. Similarly, we can extract pattern-based hyper-truss structures from hypergraphs by analyzing hyper-triangles of various patterns. This approach will improve community detection in hypergraphs and offer deeper insights into the intricate interconnections within these complex data structures.

\section{Acknowledgment}
Kai Wang is the corresponding author. Kai Wang is supported by NSFC 62302294 and U2241211. Wenjie Zhang is supported by ARC FT210100303. Ruijia Wu is supported by NSFC 12301382. Xuemin Lin is supported by NSFC U2241211. 

\newpage




\bibliographystyle{ACM-Reference-Format}
\bibliography{sample}

\end{document}